\documentclass[sigconf]{acmart}
\acmConference[ISSTA 2023]{ACM SIGSOFT International Symposium on Software Testing and Analysis}{17-21 July, 2023}{Seattle, USA}

\usepackage{array}
\usepackage{xspace}
\newcommand{\code}[1]{\texttt{#1}}
\newcommand{\name}[0]{Fluffy}

\newcolumntype{P}[1]{>{\centering\arraybackslash}p{#1}}
\newcolumntype{R}[1]{>{\raggedleft\arraybackslash}p{#1}}

\newcommand{\BinaryClassification}{Binary Classification\xspace}
\newcommand{\SinkPrediction}{Sink Prediction\xspace}
\newcommand{\NoveltyDetection}{Novelty Detection\xspace}
\newcommand{\Codex}{Codex Language Model\xspace}
\newcommand{\Frequency}{Frequency Counting\xspace}
\newcommand{\CodeQL}{Regular Expressions\xspace}

\newcommand{\BinaryClassificationShort}{Binary Class.\xspace}
\newcommand{\SinkPredictionShort}{Sink Pred.\xspace}
\newcommand{\NoveltyDetectionShort}{Novelty Det.\xspace}
\newcommand{\CodexShort}{Codex\xspace}
\newcommand{\FrequencyShort}{Frequency\xspace}
\newcommand{\CodeQLShort}{Regexps\xspace}

\begin{document}

\title{Beware of the Unexpected: Bimodal Taint Analysis}

\author{Yiu Wai Chow}
\affiliation{%
  \institution{University of Stuttgart}
  \city{Stuttgart}
  \country{Germany}}
\email{victorcwai@gmail.com}

\author{Max Sch\"afer}
\affiliation{%
  \institution{GitHub}
  \city{Oxford}
  \country{UK}
}
\email{max-schaefer@github.com}

\author{Michael Pradel}
\orcid{0000-0003-1623-498X}
\affiliation{%
  \institution{University of Stuttgart}
  \city{Stuttgart}
  \country{Germany}}
\email{michael@binaervarianz.de}

\begin{abstract}
    Static analysis is a powerful tool for detecting security vulnerabilities
    and other programming problems. Global taint tracking, in particular, can
    spot vulnerabilities arising from complicated data flow across multiple
    functions. However, precisely identifying which flows are problematic is
    challenging, and sometimes depends on factors beyond the reach of pure
    program analysis, such as conventions and informal knowledge. For example,
    learning that a parameter \code{name} of an API function \code{locale} ends
    up in a file path is surprising and potentially problematic. In contrast, it
    would be completely unsurprising to find that a parameter \code{command}
    passed to an API function \code{execaCommand} is eventually interpreted as
    part of an operating-system command.
    This paper presents \name{}, a bimodal taint analysis that combines static
    analysis, which reasons about data flow, with machine learning, which
    probabilistically determines which flows are potentially problematic. The
    key idea is to let machine learning models predict from natural language
    information involved in a taint flow, such as API names, whether the flow is
    \emph{expected} or \emph{unexpected}, and to inform developers only about
    the latter. We present a general framework and instantiate it with four
    learned models, which offer different trade-offs between the need to
    annotate training data and the accuracy of predictions. We implement \name{}
    on top of the CodeQL analysis framework and apply it to 250K JavaScript
    projects. Evaluating on five common vulnerability types, we find that
    \name{} achieves an F1 score of 0.85 or more on four of them across a
    variety of datasets.
\end{abstract}

\maketitle

\section{Introduction}
Taint analysis is a powerful technique for detecting various kinds of
programming mistakes, including both security vulnerabilities and other kinds of
bugs. A taint analysis tracks the flow of information from a \emph{source} to a
\emph{sink} and reports a warning if the existence of this flow violates some
pre-defined policy. For example, such a policy might specify that data from an
HTTP request sent by an unknown and potentially malicious user (the source) must
not flow into an API that executes strings as operating-system commands (the
sink) to prevent remote code execution, a type of vulnerability known as
\emph{command injection}. As another example, a policy might state that
unencrypted passwords or other secrets must not flow into logging statements
to prevent accidentally leaking them through \emph{clear-text logging}. Taint
analysis is an integral part of popular static analysis solutions, such as Facebook Infer
or GitHub Code Scanning, that regularly scan large amounts of proprietary and open-source
software.

A lot of research and engineering has gone into developing precise and scalable
algorithms for inter-procedural taint tracking for various languages,
ensuring that modern taint analyses can scale to even the largest of code bases.
However, in the end a taint analysis is only as good as the policies it enforces:
If a source or sink is missing, the analysis will not report problematic flows
involving it, leading to false negatives and hence missed vulnerabilities.
If, on the other hand, a source or sink is included that should not be, the analysis will report false positives, leading to unnecessary work and frustration for developers and security researchers.

In some cases, such as the command injection example above, suitable sources and
sinks can be identified by carefully examining the APIs of the analyzed libraries, which is a
well-defined if perhaps tedious process that can cover only a limited set of APIs due to the high human effort.
For clear-text logging, on the other hand, the solution is not so obvious.
While it is easy to identify the sinks (calls to logging libraries), it is much
less clear what constitutes a source, since the way sensitive data are stored and
handled is usually highly application-specific.

As another example, consider the problem of detecting ``unhygienic'' APIs, that
is, library APIs that use data from a client application in a security-sensitive
context without clearly indicating to the client that this is the case (for
example by appropriate naming or documentation).
Depending on whether the
security-sensitive use is intentional or not, such APIs either suffer from a
vulnerability or provide insufficient documentation, both of which put client
applications at risk. As an example, the API of the JavaScript library
\texttt{moment} for handling times and dates contains a \texttt{locale} function
for setting the locale used for pretty-printing purposes. Before version 2.29.2,
this function treats its parameter \texttt{name} (the name of the locale to
use) unsafely, leading to remote code execution if untrusted data is passed to
it.%
\footnote{\href{https://github.com/advisories/GHSA-8hfj-j24r-96c4}{CVE-2022-24785}.}
Nothing about the name of the function or the parameter suggests that the parameter directly impacts an executed command, and indeed, this behavior was not intentional but a consequence of sloppy coding practices.
A client relying on API names and documentation, might well
pass data obtained from an untrusted third party to the API
and thereby open themselves up to a vulnerability--as some, indeed, did
before the \texttt{moment} developers fixed the problem.\footnote{For example, see advisory \href{https://github.com/TryGhost/Ghost/security/advisories/GHSA-7v28-g2pq-ggg8}{GHSA-7v28-g2pq-ggg8}.}%

Identifying such unhygienic APIs seems to be a natural fit for taint analysis,
but while it is easy to define the sinks (they are identical to the sinks for
command injection), the set of sources is harder to pin down. Just treating all
API entry points as sources would not do: For example, the library
\texttt{execa}, which provides improved cross-platform support for spawning
child processes on Node.js, exports a function \texttt{execaCommand} with a
parameter \texttt{command} that flows into a command-injection sink.
However, flagging this flow as a vulnerability makes no sense, since it is precisely the intended functionality.

The common denominator in both of these examples is that the flows we are
looking for are in some sense \emph{unexpected} in that they are at odds with
the intentions and expectations of developers: In a production code base, one
would not expect to see sensitive information being logged in clear text, and
likewise one would not expect a library to use data in a security-sensitive
context without clearly indicating to the client that this is the case.

To address the problem of detecting such unexpected taint flows, 
we present \name{} (``\underline{Fl}agging \underline{u}nexpected \underline{f}lows \underline{f}or better securit\underline{y}'') a \emph{bimodal taint analysis} that exploits the fact that source code conveys
meaning both through the programming language semantics and through natural
language information embedded in code, e.g., in the form of identifier
names~\cite{Hindle2012,Allamanis2018,NeuralSoftwareAnalysis}. \name{} allows for
implementing security policies where the set of sources is partially determined using
natural language information.%
\footnote{The case of sinks is symmetric, but in this paper we focus on sources only.}

The approach consists of two steps. First, we run an off-the-shelf, monomodal taint
analysis that \emph{overapproximates} the set of sources so it can be
captured purely in terms of code constructs without any natural language
reasoning. For example, for clear-text logging we consider \emph{all} variables to
be sources, and for unhygienic APIs all parameters. Secondly, the
\emph{candidate flows} resulting from the first step are filtered by a
machine learning model that predicts whether the source in a candidate flow is, in fact, a true source.
Our approach combines the power of logic-based static taint analysis, which
reasons about data flow and control flow dependencies, and machine learning
models, which probabilistically ``understand'' the meaning of natural language
information. 

The \name{} approach is a general framework with four concrete
instantiations that use different ways of formulating the prediction task, build
on different machine learning models, and impose different demands for manually
labeled data. The four instantiations are
(i) a neural, binary classifier trained on manually labeled examples, 
(ii) a neural model trained on millions of taint flows extracted automatically via static analysis,
(iii) a novelty detection technique based on a one-class support vector machine, and
(iv) a technique that queries a large language model (Codex~\cite{Chen2021}) via few-shot learning.

Our work relates to combinations of program analysis and ML
models for other analysis problems, such as call graph
pruning~\cite{Utture2022,Le-Cong2022} and filtering null dereference
warnings~\cite{Kharkar2022}. \name{} differs by addressing the problem
of identifying unexpected taint flows. Another related line of work is a
purely neural taint analysis~\cite{She2019}, which trains a neural model to
``emulate'' a dynamic taint analysis more efficiently than an actual dynamic
analysis would be. In contrast, our work leverages the complementary power of static analysis and machine learning.
Finally, \name{} also relates to prior work on automatically annotating APIs as
sources and sinks using a trained model~\cite{susi}. Their work relies on a set
of hand-coded features, such as whether a method returns a specific type, and it
makes predictions about source and sink APIs. In contrast, we exploit
pre-trained neural models that avoid manual feature engineering and propose
models that make predictions about specific flows between already identified
sources and sinks. Overall, our work is the first to present a bimodal analysis
for the problem of taint tracking.

We implement the approach in the CodeQL analysis framework~\cite{codeql} and
apply it to 250k JavaScript and TypeScript projects. Our evaluation considers
five taint analyses aimed at detecting integrity problems, such as command
injections, and confidentiality problems, such as clear-text logging. We find
\name{} to be effective at determining unexpected taint flows, with 81\%--97\%
precision, 80\%--100\% recall, and 76\%--97\% F1-score, depending on the
analysis. To assess the usefulness of the approach, we apply \name{} to 131
confirmed vulnerabilities from the past, of which it successfully detects 117. 
Finally, we report 16 newly detected vulnerabilities to the respective developers,
who have so far confirmed eight of them.

In summary, this paper contributes the following:
\begin{itemize}
    \item A bimodal taint analysis, combining static analysis with machine learning to identify problematic flows.
    \item A general framework with four instantiations, offering trade-offs between labeling effort and prediction accuracy.
    \item An integration into CodeQL and empirical evidence that the approach is effective and useful in practice.
\end{itemize}

To foster future work, our implementation and experimental results are publicly available.\footnote{\url{https://figshare.com/s/1ab456424bfb5a2ead5e}}

\section{Background: Taint Analysis with CodeQL}
\label{sec:background}

CodeQL~\cite{codeql} is an open-source static program analysis system, which powers LGTM.com and GitHub
CodeScanning. Following the code-as-data paradigm, code is represented as
relational data and analyses are expressed as queries written in QL~\cite{ql},
an object-oriented extension of Datalog. CodeQL offers extensive standard
libraries for performing typical analysis tasks, including a framework for
global taint analysis that underpins a suite of individual analyses for
finding common vulnerabilities, such as command injection or path traversal.

Security policies are also expressed in QL, typically identifying uses of
particular APIs as sinks and sources of taint. In the case of JavaScript, for
instance, the CodeQL query for identifying command-injection vulnerabilities%
\footnote{\url{https://github.com/github/codeql/blob/main/javascript/ql/src/Security/CWE-078/CommandInjection.ql}.}
identifies various sources of potentially untrusted user input, such as HTTP
request parameters accessed through the \texttt{express} npm package, as sources,
and calls to command execution APIs, such as those in the Node.js
\texttt{child\_process} module, as sinks. The path-traversal query%
\footnote{\url{https://github.com/github/codeql/blob/main/javascript/ql/src/Security/CWE-022/TaintedPath.ql}.}
shares the same set of sources, but uses file-system operations as its sinks.

Both of these are examples of \emph{integrity violations} whereby untrusted data
is used in a security-sensitive context. Other examples of such violations also
covered by CodeQL queries include code injection,%
\footnote{\url{https://github.com/github/codeql/blob/main/javascript/ql/src/Security/CWE-094/CodeInjection.ql}.}
where untrusted data is used to construct code to be executed, and reflected
cross-site scripting (``reflected XSS'' for short),%
\footnote{\url{https://github.com/github/codeql/blob/main/javascript/ql/src/Security/CWE-079/ReflectedXss.ql}.}
where untrusted data is embedded in an HTTP response.

As mentioned above, some analyses are not amenable to this kind of
formulation. One example is the problem of finding clear-text logging of
sensitive information, an example of a \emph{confidentiality violation}
in which security-sensitive data is exposed in an untrusted context. The CodeQL
query for finding clear-text logging%
\footnote{\url{https://github.com/github/codeql/blob/main/javascript/ql/src/Security/CWE-312/CleartextLogging.ql}.}
employs hand-crafted regular expressions to identify variables and properties
whose names suggest they might contain sensitive data, which are then treated as
sources for the analysis. As we will demonstrate below, this is error-prone and
a manual update to the libraries is required every time a spurious or missing
source is discovered.

As another example, the CodeQL query \textit{Unsafe shell command constructed from
library input}%
\footnote{\url{https://github.com/github/codeql/blob/main/javascript/ql/src/Security/CWE-078/UnsafeShellCommandConstruction.ql}.}
attempts to identify unhygienic library APIs by detecting taint
flow from API parameters to shell commands, excluding parameters whose name
suggests they are meant to be commands. Again, a regular expression is used for
this purpose, with the same drawbacks as above.

Recently, CodeQL has been extended with experimental support for leveraging
machine-learning techniques to discover new sources and sinks that are not
covered by the standard security policies~\cite{atm}. While this helps with
identifying sources and sinks in rarely-used or proprietary APIs, it is not
clear whether it offers any advantages in situations where natural language
information is needed since it is currently only available for four queries,
none of which fall into this category.

\section{Approach}

\begin{figure}
    \includegraphics[width=.8\linewidth]{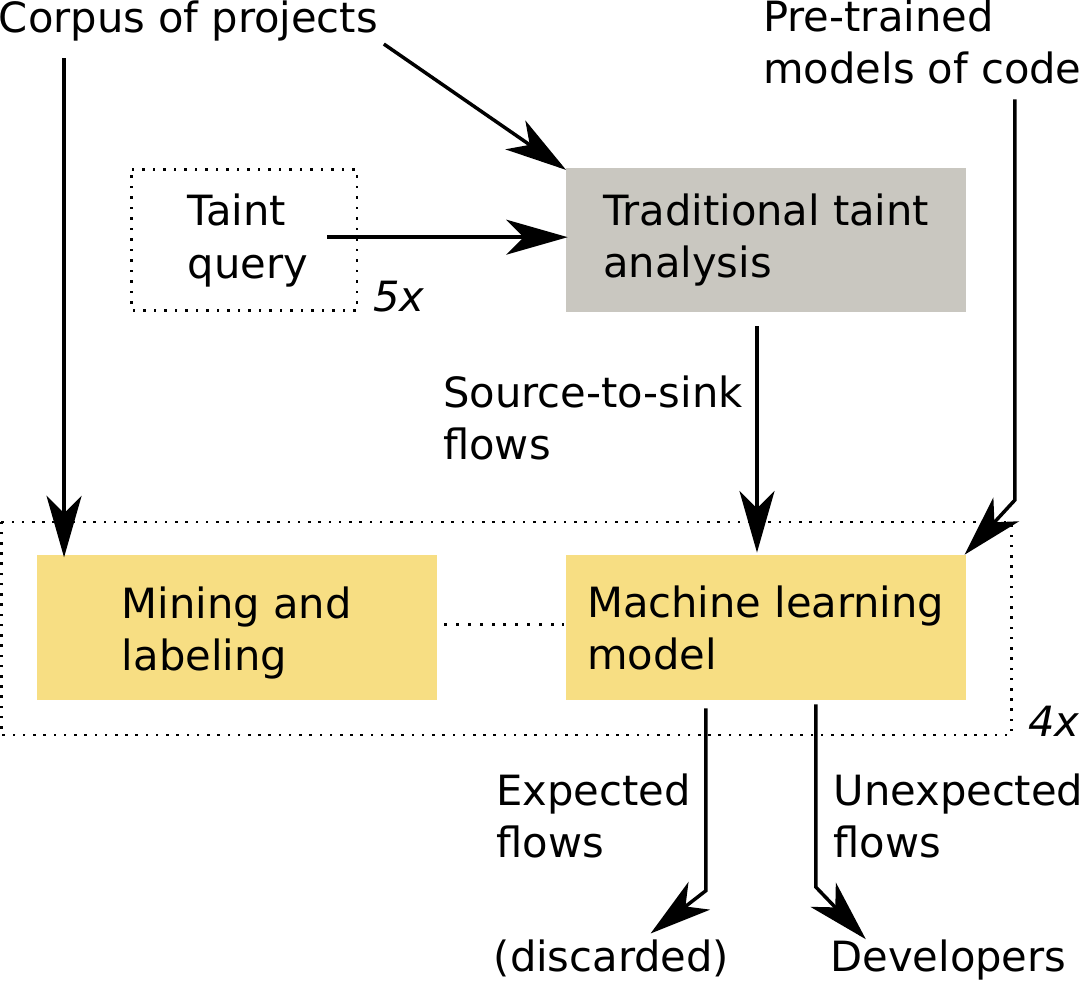}
    \caption{Overview of the approach. The yellow components are the contributions of this paper.}
    \label{fig:overview}
\end{figure}

This section presents \name{}, our bimodal taint analysis, in more detail.
After defining the problem addressed by the approach (Section~\ref{sec:pb}), Section~\ref{sec:overview} gives an overview of our general framework.
The remaining subsections then present four instantiations of our framework, which explore different trade-offs in the design space.

\subsection{Problem Statement}
\label{sec:pb}

The input to \name{} is a taint-style security policy $\mathcal{P}$ and a code
base~$C$ to check against that policy. The security policy specifies a set of
problematic flows for $C$ as a set of source-sink pairs $(l_{src}, l_{sink})$
for which no data should propagate from the source location~$l_{src}$ to the
sink location~$l_{sink}$ in $C$.

This specification consists of two parts $\mathcal{P}_F$ and $\mathcal{P}_U$.
The first part $\mathcal{P}_F$ is \emph{precisely specified but
overapproximate}, determining a set~$F$ of potentially problematic flows that
can be computed purely in terms of code structure, for instance by a standard
taint analysis. The second part $\mathcal{P}_U$ is \emph{imprecisely specified}, describing a set $U$ of unexpected flows. Whether a flow is
unexpected is an inherently fuzzy problem, because the expectations of
developers depend on common coding conventions and informal information, such as
identifier names and API documentation.

The problem addressed in this paper is to identify all flows in a given code
base that fulfill both parts of the security policy, i.e., to find the set $\{
(l_{src}, l_{sink}) \mid l_{src} \in C, l_{sink} \in C, (l_{src}, l_{sink}) \in
F \cap U \}$ of actually problematic flows to show to the developer.

\begin{table*}
    \caption{Four instantiations of our general framework.}
    \label{tab:instantiations}
    \begin{tabular}{@{}lp{.36\textwidth}p{.41\textwidth}@{}}
        \toprule
        Approach & Training data                                                               & Model                                                                      \\
        \midrule
        \BinaryClassification  & Hundreds of manually labeled flows & Neural binary classifier, predicts if a flow is expected \\
        \SinkPrediction & Millions of unlabeled flows & Neural model, predicts the most likely sink                                        \\
        \NoveltyDetection  & Fewer than ten relevant terms for each sink type & One-class SVM, identifies unusual flows \\
        \Codex & Ten examples given in the prompt & Generative language model, predicts if a flow is expected \\
        \bottomrule
    \end{tabular}
\end{table*}

\subsection{Overview}
\label{sec:overview}

Motivated by the fuzziness of the problem, we address it with learning-based techniques, which have been shown to effectively capture inherently fuzzy information~\cite{NeuralSoftwareAnalysis}.
Figure~\ref{fig:overview} gives an overview of the approach.
In line with the two-part specification of the problem, the approach consists of two main steps.
Given a project to analyze, the first step is to apply a traditional taint analysis.
We build upon CodeQL, which offers a static code analysis engine and a suite of standard taint analyses as described in Section~\ref{sec:background}.
The taint analysis $T\colon C \mapsto F$ identifies all flows between source-sink pairs in the given code base~$C$ that are in $F$.
In the second step, the source-sink flows detected by the taint analysis are then given to a machine learning model $M\colon F \mapsto U$ that predicts which flows in $F$ are likely unexpected by the developers, and hence should be reported to the developers.
The machine learning model builds on existing, pre-trained models, e.g., to embed code tokens into a continuous vector, and is supported by a mining and labeling component, which provides training data to learn from.

The overall framework described in Figure~\ref{fig:overview} can be instantiated in several ways.
One dimension is the taint analysis to use; we focus on five CodeQL analyses for JavaScript: code injection, command injection, reflected XSS, path traversal, and clear-text logging.
The other dimension is the machine learning model to use; we present four models that explore different trade-offs in the design space.
Table~\ref{tab:instantiations} gives an overview of the four machine learning models, which are described in detail in Sections~\ref{sec:nlp} and~\ref{sec:models}.

\subsection{Gathering and Representing Natural Language Information}
\label{sec:nlp}

Running the queries described above against a code base yields a set $F$ of flows, i.e., source-sink pairs where data from the source propagates to the sink.
To determine whether a flow is likely unexpected by developers, \name{} exploits natural language information associated with the flow.
Specifically, the approach gathers the following information about each flow:
\begin{itemize}
\item The identifier name $n_s\in N$ of the source, which is the parameter $p$ of an API function $f$ for the integrity queries and a source variable $v$ for the confidentiality query.
\item Additionally, for the integrity queries, the name $n_f \in N_{\mathit{fct}}\subseteq N$ of $f$ and any documentation $d \in D$ for $p$.
\end{itemize}

To allow a machine learning model to effectively reason about the extracted natural language information, we embed it into a continuous vector representation.
For this purpose, \name{} uses a pre-trained embedding function $e: N \rightarrow \mathbb{R}^k$ that maps a natural language string in $N$ to a $k$-dimensional vector in such a way that semantically similar strings have similar vector representations.
As the embedding function, we use VarCLR~\cite{Chen2022}, which we select for two reasons.
First, VarCLR has been shown on the IdBench benchmark~\cite{icse2021} to be the state of the art for the task of mapping identifier names to continuous vectors in a way that preserves semantic relatedness and similarity.
Second, while comparing VarCLR against other pre-trained embeddings, e.g., FastText~\cite{Bojanowski2017}, in preliminary experiments, we found VarCLR to produce the best overall results.

\subsection{Predicting Unexpected Flows}
\label{sec:models}

Based on the flows $F$ extracted by traditional taint analysis and the natural language information associated with each flow, the core step of \name{} is to determine the subset $U \cap F$ of flows that are unexpected and hence should be reported to developers.
The following presents four machine learning-based techniques for this purpose, which differ in the kind and amount of labeled training data they require, the kind of model they build on, and (as shown in Section~\ref{sec:evaluation}) their ability to effectively identify unexpected flows.

\subsubsection{Approach 1: \BinaryClassification}

We train a neural model $M$ that predicts the probability that a given flow is unexpected.
As its input, the model receives embeddings of the identifier name of the source, and if available the other natural language information associated with the flow.
That is, the model is the following learned function:
$$M\colon N \times N_{\mathit{fct}} \times D \rightarrow \{ \mathit{Expected}, \mathit{Unexpected} \}$$

The model is implemented using a standard feed-forward neural network with two hidden layers and the softmax function applied to the output layer, which we train with cross-entropy loss.
The names $n_s$ and $n_f$ are embedded using the embedding function~$e$.
For the documentation $d$, we embed each token using a jointly trained embedding layer and encode the sequence of embedded tokens using an LSTM-based, bidirectional recurrent neural network.
We train a separate model for each of the five taint queries, using hundreds of manually labeled flows (details in Section~\ref{sec:evaluation}).

\subsubsection{Approach 2: \SinkPrediction}

The \BinaryClassification approach requires a significant amount of manually labeled training data.
To avoid this human effort, to following approach is trained only on automatically mined training data.
The basic idea is to train a neural classification model that predicts for a given source what sink it is supposed to flow into. %

For the integrity queries, this corresponds to the function
$$M\colon N \times N_{\mathit{fct}} \times D \rightarrow \{ \mathit{CmdInj}, \mathit{CodeInj}, \mathit{XSS}, \mathit{PathTrav}, \mathit{None} \}$$
where the first four output classes correspond to the sinks of the four integrity queries, and $\mathit{None}$ means that a source location does not propagate to any of the any four sinks.
Likewise, for the confidentiality query, the model is the following learned function:
$$M\colon N \rightarrow \{ \mathit{Logging}, \mathit{None} \}$$
where $\mathit{Logging}$ means that it is unproblematic for the source contents to be logged, while $\mathit{None}$ means that it should not be logged.

The model uses the same architecture as for the \BinaryClassification, but with an output layer that has the appropriate length for the classification problem.
Training data is collected using CodeQL, as explained in Section~\ref{sec:data-collection}, without any need for manual labeling.

Once the model has been trained, \name{} queries it with previously unseen flows to determine whether the flow is unexpected.
If the flow involves a sink type $s\neq\mathit{None}$ and the model predicts $s$ as the most likely sink, then we consider the flow to be expected.
The rationale is that if sources with particular names commonly flow this sink type, then developers are likely aware of this.
Otherwise, if the model predicts some other sink $s' \neq s$ (where $s'$ may be $\mathit{None}$), we consider the flow to be unexpected and report it.

\subsubsection{Approach 3: \NoveltyDetection}

The previous two approaches require training data labeled to belong to different classes.
Our third approach instead uses a novelty detection technique trained only on examples of one class.
We use a one-class support vector machine (OC-SVM)~\cite{scholkopf2001estimating} that predicts whether a previously unseen example belongs to this class or is ``novel''.

\begin{table}
    \caption{Seed names used for \NoveltyDetection.}
    \label{tab:seed names}
    \begin{tabular}{@{}p{8em}p{17em}@{}}
        \toprule
            Sink type & Seed names \\
        \midrule
            \multicolumn{2}{@{}l@{}}{\emph{Integrity:}} \\
        \midrule
            Command injection & \texttt{execute}, \texttt{command} \\
            Code injection & \texttt{eval}, \texttt{execute}, \texttt{compile}, \texttt{render}, \texttt{callback}, \texttt{function}, \texttt{fn} \\
            Reflected XSS & \texttt{sent}, \texttt{content} \\
            Path traversal & \texttt{file}, \texttt{directory}, \texttt{path}, \texttt{cwd}, \texttt{source}, \texttt{input} \\
        \midrule
            \multicolumn{2}{@{}l@{}}{\emph{Confidentiality:}} \\
        \midrule
            Clear-text logging & \texttt{authkey}, \texttt{password}, \texttt{passcode}, \texttt{passphrase} \\
        \bottomrule
    \end{tabular}
\end{table}

For each sink kind, we train an OC-SVM model with examples of expected ``seed'' names,
as shown in Table~\ref{tab:seed names}, which we select based on our understanding of the domain.
For the integrity queries these names indicate data that we would expect to flow to the specific sink,
while for the confidentiality query they indicate sensitive data that is \emph{not} expected to flow to the sink.
As the set of names is small (less than ten examples per sink kind), obtaining this training data imposes relatively little effort.
To train the model, we embed the natural language information using the pre-trained embedding~$e$. 
Once trained, we let the model predict whether a previously unseen flow is unexpected,
again embedding the natural language information using $e$.

\subsubsection{Approach 4: \Codex}
\label{sec:codex}

Motivated by the impressive results obtained with large language models like Codex~\cite{Chen2021}, for different software engineering tasks~\cite{codexStudy2022,Jain2022,Kharkar2022,Joshi2022}, we also implement a Codex-based approach.
To query the language model, we design a prompt that provides information about the flow and ends with a comment that we ask the model to complete.
Following a few-shot approach~\cite{Brown2020}, the prompt includes ten examples of the task we want the model to perform and ends with an unfinished eleventh example for it to complete.

For the integrity queries, the prompt consists of the following:
\begin{itemize}
    \item The signature of the API function and, if available, the doc comment associated with it.
    \item A JavaScript comment with the sentence ``In the above function $f$, the parameter $p$ flows into the $s$ sink ($e$), which is $c$'', where
    \begin{itemize}
        \item $f$ is the name of the function,
        \item $p$ is the name of the formal parameter,
        \item $s$ is an abbreviated name of the query, e.g., ``CommandInjection'',
        \item $e$ is a brief explanation of the query, e.g., ``uncontrolled data used in a path expression'', and
        \item $c$ is either ``expected'' or ``unexpected''.
    \end{itemize}
\end{itemize}
For the confidentiality query, the prompt consists of:
\begin{itemize}
    \item A stub function $f$ calling \texttt{console.log} on its parameter $p$.
	\item A JavaScript comment with the sentence ``In the above function $f$, the parameter $p$ is being logged, which likely exposes $c$ data'', where
        $c$ is either ``sensitive'' or ``insensitive''.
\end{itemize}

The word in $c$ is provided for the ten few-shot examples but left undefined for the eleventh example.
The model then completes the prompt by predicting the missing word.
For the integrity queries, we check whether ``expected'' or ``unexpected'' is the more likely completion according to the model, and report flows to the developers only when the model predicts them as unexpected.
Likewise, for the confidentiality query, we report a flow to developers only when the model considers completing the prompt with ``sensitive data'' as more likely than completing it with ``insensitive data''.

\section{Evaluation}
\label{sec:evaluation}

To assess the efficacy and practicality of our approach, we pose ourselves the following research questions:

\begin{description}
    \item[RQ1] How effective is \name{} at identifying unexpected flows?
    \item[RQ2] How effective is \name{} at finding vulnerabilities?
    \item[RQ3] What is the trade-off between labeling effort and prediction accuracy?
    \item[RQ4] How well does \name{} scale to large code bases?
\end{description}

We now describe our evaluation setup, methodology, and results, and discuss how
they answer our research questions.

\subsection{Experimental Setup}

\subsubsection{Data Collection}
\label{sec:data-collection}
\begin{sloppypar}
We collect three datasets for our evaluation. The first two
(\textit{param-sink flows} and \textit{logging flows}) are compiled by running
special-purpose CodeQL queries across all JavaScript/TypeScript projects on
LGTM.com (around 250,000 at the time of writing), while the third one is derived
from \textit{SecBench.js}~\cite{secbench}, a corpus of real-world
vulnerabilities in server-side JavaScript.
\end{sloppypar}

\paragraph{Param-sink flows.}
The first dataset concerns unhygienic APIs. We want to collect examples of
flows from API parameters to known sinks. For purposes of this dataset, an API
parameter is a parameter of a function exported by an npm package, or (by a
slight abuse of terminology) a property of such a parameter. Our sinks are
collected from four CodeQL security analyses that identify four common types of
integrity violations: code injection, command injection, reflected XSS, and path
traversal.

For each kind of integrity violation, we write a custom CodeQL query using the standard taint-tracking framework.
The queries collect flow tuples of the form $(p, s)$, where $p$ is an API parameter and $s$
is one of the four types of sinks mentioned, such that taint flows from $p$ into a sink of type $s$, or the special type \textit{None} if no flow from $p$
to any known sink is found.\footnote{Note this means that we can have two flows $(p, s)$ and $(p, s')$ for
the same parameter $p$ with different sink types $s$ and $s'$, but only if $s$
and $s'$ are not \textit{None}.}
Apart from the name of the API parameter, we also collect its doc comment and
the doc comment of the enclosing function, if any. Finally, we filter out
flows where the parameter name has less than two characters, since
single-character names are unlikely to convey much semantic information to
either a human or a model.

\begin{sloppypar}
Overall, the query found 3,245,860 flows on 61,123 projects, where 3,228,034 are
flows to \textit{None}; 1,123 to code-injection sinks; 1,498 to command-injection
sinks; 70 to reflected-XSS sinks; and 15,135 to path-traversal sinks. Note the
grave imbalance of sink types, which is a consequence of the selection of projects on LGTM.com, and hence not under our control.
Filtering out single-character names removed 233,091 flows, i.e., 6.7\% of all flows before filtering.
\end{sloppypar}

\paragraph{Logging flows.}
The second dataset concerns clear-text logging. We again write a custom CodeQL
taint-tracking query, this time looking for flow from a variable or property
$v$ into a call to a logging function, as determined by existing API modeling in
the CodeQL standard libraries. Apart from the name of $v$, we also record
whether the clear-text logging analysis shipping with CodeQL considers $v$ to be
potentially sensitive. As above, we filter out single-character names, which account for 8.9\% of all flows,
yielding 4,535,851 flows in 112,765 projects.

\paragraph{SecBench.js.}
\begin{sloppypar}
For our final dataset, we examine all code-injection, command-injection, and
path-injection vulnerabilities in the SecBench.js corpus to find the ones caused
by unhygienic APIs, of which there are 33, 101, and 1, respectively after
filtering.\footnote{We exclude one code-injection vulnerability and two
command-injection vulnerabilities from consideration since the affected npm
package is no longer available, making it impossible to determine the parameters
involved. We also exclude one further command injection vulnerability for which
details are no longer available since the corresponding advisory has been
withdrawn.}
\end{sloppypar}

\begin{table*}[t]
	\begin{center}
		\caption{Comparison of the different approaches.}
		\setlength{\tabcolsep}{16pt}
		\label{table:method_summary}
        \begin{tabular}{@{}lllll@{}} 
			\toprule
			Approach & Training data & Number of models & Needs threshold & Computation of F1 score \\
			\midrule
			\SinkPrediction & Param-sink flows & One & Yes & From precision-recall curve \\
			\NoveltyDetection & Seed names & One per sink type & Yes & From precision-recall curve \\
			\BinaryClassification & Balanced set & One per sink type & No & From k-fold cross-validation\\
			\CodexShort & Balanced set & One & No & Direct \\
			\FrequencyShort & Param-sink flows & One & Yes & From precision-recall curve \\
			\bottomrule
		\end{tabular}
	\end{center}
\end{table*}

\begin{table}
	\centering
	\caption{Ground truth labels.}
	\label{table:ground_truth}
	\setlength{\tabcolsep}{4pt}
	\begin{tabular}{@{}lrrrr@{}}
		\toprule
		& \multicolumn{2}{c}{Random Set} & \multicolumn{2}{c}{Balanced Set} \\ 
		\cmidrule(lr){2-5} 
		\emph{Integrity:} & Unexpected & Total & Unexpected & Total \\ 
		\midrule
		Code injection & 16 & 27 & 113 & 340 \\
		Command injection & 15 & 29 & 144 & 168 \\
		Reflected XSS & 19 & 28 & 29 & 46 \\
		Path traversal & 8 & 188 & 105 & 504 \\
        \toprule
		\emph{Confidentiality:} & & & Unexpected & Total \\        
		\midrule
		Logging sensitive data & & & 245 & 340 \\
		\bottomrule
	\end{tabular}%
\end{table}

\subsubsection{Non-Neural Baselines}
In addition to the four neural network-based instantiations of \name{}, we also consider two
non-neural baseline approaches to compare against.
The first baseline, called \Frequency, is a statistical approach based on determining the frequency with which a
given parameter name is observed to flow into a particular sink type in our
param-sink flows dataset. If this frequency is below a given threshold, we
consider this sink type to be unexpected for the parameter. Note that no
semantic reasoning about names is involved and only exact name matches are
considered. Due to the composition of the datasets this approach only works for
the integrity violations since the logging-flows dataset considers only a
single sink type.

The second baseline, called \CodeQL, is for the logging flows.
It is an existing CodeQL query (Section~\ref{sec:background}) that uses regular expressions to flag likely problematic flows,
first applying two regular expressions to identify names that may indicate sensitive content such as \code{passcode} and \code{authkey}.%
\footnote{\url{https://github.com/github/codeql/blob/499f20f6e8a3a91e394c30e05a340fe10b9ecec7/javascript/ql/lib/semmle/javascript/security/internal/SensitiveDataHeuristics.qll\#L69}}
Names matching either of these regular expressions are then checked against a third regular
expression that filters out names suggesting that the value has been encrypted or hashed.%
\footnote{\url{https://github.com/github/codeql/blob/499f20f6e8a3a91e394c30e05a340fe10b9ecec7/javascript/ql/lib/semmle/javascript/security/internal/SensitiveDataHeuristics.qll\#L104}}

We also considered a direct comparison with CodeQL queries that search for confidentiality violations and that existed prior to this work (Section~\ref{sec:background}).
However, these queries do not answer the same question as \name{}.
The preexisting queries find end-to-end flows from sources of tainted data to sensitive sinks, whereas \name{} works on flows from API parameters to sensitive sinks.
The preexisting queries would not consider these API parameters to be sources (and conversely, \name{} does not consider flows from non-parameter sources), so the two cannot be compared directly.

\subsubsection{Data Labeling}
To obtain ground truth for our evaluation, we label a subset of the collected
data to indicate whether the flow may cause or contribute to a vulnerability.

For the param-sink dataset, the authors manually label 1,058 flows. To check
the reliability of our labeling, we send out a survey to four program
analysis engineers with experience in security-related analyses.
The survey asks the analysis experts to label 30 randomly selected flows each, based
on the parameter name, function name, package name, sink, package description,
parameter documentation, and function documentation.
Specifically, the question we ask is: ``As a client of the package, would you expect data passed into this parameter to flow into the sink in question without further sanitization?''
Using Krippendorff's alpha to compute the inter-rater agreement, the labels we assign and those given by the analysis experts have an agreement score of 0.74.
That is, the analysis experts by and large agree with the labels in our ground truth. 

The total set of labeled flows consists of two overlapping subsets. The
first is the \textit{random set}, which consists of 272 randomly selected flows.
While this set is unbiased, it does not contain many unexpected flows since
most flows observed in real-world code are unproblematic. Hence, we extend the random set into a second set, which we call the \textit{balanced set}, additionally including flows selected from results of the \Frequency approach.

For the logging dataset, we manually label a subset of 340 flows.
Table~\ref{table:ground_truth} shows the number of total and unexpected flows
according to our labeling for each sink type.

Note that our manual labeling shows that simply flagging all flows as unexpected
would result in many false positives, demonstrating the importance of our
approach.

Finally, for the SecBench.js dataset, we review all flows, and identify one
code-injection vulnerability and seven command-injection vulnerabilities that, despite having been reported as a vulnerability, are expected according to our judgment.
Full details about these vulnerabilities are included in the supplementary materials.
To give just one example, SecBench.js includes
\href{https://github.com/advisories/GHSA-ww4j-c2rq-47q8}{CVE-2020-7784}, which
identifies the fact that the parameter \code{command} of the function
\code{exec} exported by the npm package \code{ts-process-promises} is
interpreted as a shell command. We consider this flow obvious from context and
hence not a genuine vulnerability. None of the vulnerabilities have been
fixed, and in one case the advisory has been withdrawn, suggesting that the
authors of the packages involved do not consider them to be a problem.
All remaining 127 vulnerabilities are labeled as unexpected by us.

\begin{table*}[ht]
	\centering
	\caption{Effectiveness of \name{} on the random set of param-sink flows.}
	\setlength{\tabcolsep}{4pt}
	\label{table:result_random}
	\begin{tabular}{@{}lrrr|rrr|rrr|rrr@{}}
		\toprule
		& \multicolumn{12}{c}{Random Set} \\ 
		\cmidrule{2-13} 
		& \multicolumn{3}{c|}{Code injection} & \multicolumn{3}{c|}{Command injection} & \multicolumn{3}{c|}{Reflected XSS} & \multicolumn{3}{c}{Path traversal} \\ 
		\cmidrule{2-13}
		Approach & Precision & Recall & F1-Score & Precision & Recall & F1-Score & Precision & Recall & F1-Score & Precision & Recall & F1-Score \\ 
		\midrule
		\SinkPredictionShort & 0.78 & 0.88 & 0.82 & 1.00 & 1.00 & \textbf{1.00} & 0.68 & 1.00 & 0.81 & 0.18 & 0.75 & 0.29\\
		\NoveltyDetectionShort & 0.82 & 0.88 & \textbf{0.85} & 1.00 & 1.00 & \textbf{1.00} & 0.86 & 1.00 & \textbf{0.93} & 0.36 & 0.63 & \textbf{0.45}\\
		\BinaryClassificationShort & 0.83 & 0.82 & 0.81 & 1.00 & 1.00 & \textbf{1.00} & 0.95 & 0.90 & 0.91 & 0.33 & 0.30 & 0.31\\
		\CodexShort & 0.65 & 0.69 & 0.67 & 0.63 & 1.00 & 0.77 & 0.94 & 0.89 & 0.92 & 0.08 & 0.13 & 0.10\\
		\midrule
		\FrequencyShort & 0.59 & 1.00 & 0.74 & 0.91 & 0.67 & 0.77 & 0.68 & 1.00 & 0.81 & 0.22 & 0.25 & 0.24\\
		\bottomrule
	\end{tabular}%
\end{table*}

\begin{table*}[]
	\centering
	\caption{Effectiveness of \name{} on the balanced set of param-sink flows.}
	\setlength{\tabcolsep}{4pt}
	\label{table:result_whole}
	\begin{tabular}{@{}lrrr|rrr|rrr|rrr@{}}
		\toprule
		& \multicolumn{12}{c}{Balanced Set} \\ 
		\cmidrule{2-13} 
		& \multicolumn{3}{c|}{Code injection} & \multicolumn{3}{c|}{Command injection} & \multicolumn{3}{c|}{Reflected XSS} & \multicolumn{3}{c}{Path traversal} \\ 
		\cmidrule{2-13}
		Approach & Precision & Recall & F1-Score & Precision & Recall & F1-Score & Precision & Recall & F1-Score & Precision & Recall & F1-Score \\ 
		\midrule
		\SinkPredictionShort & 0.36 & 0.98 & 0.53 & 0.94 & 0.99 & \textbf{0.97} & 0.63 & 1.00 & 0.77 & 0.53 & 0.64 & 0.58\\
		\NoveltyDetectionShort & 0.74 & 0.88 & 0.80 & 0.92 & 0.97 & 0.95 & 0.88 & 1.00 & \textbf{0.94} & 0.50 & 0.80 & 0.62\\
		\BinaryClassificationShort & 0.90 & 0.88 & \textbf{0.88} & 0.97 & 0.97 & \textbf{0.97} & 0.96 & 0.94 & \textbf{0.94} & 0.81 & 0.73 & \textbf{0.76}\\
		\CodexShort & 0.76 & 0.76 & 0.76 & 0.90 & 0.97 & 0.94 & 0.90 & 0.93 & 0.92 & 0.62 & 0.42 & 0.50\\
		\midrule
		\FrequencyShort & 0.33 & 1.00 & 0.50 & 0.93 & 0.92 & 0.93 & 0.63 & 1.00 & 0.77 & 0.42 & 0.63 & 0.51\\
		\bottomrule
	\end{tabular}%
\end{table*}

\subsubsection{Model Hyperparameters}
All our models operate on vectors with 768 dimensions, corresponding to the
output dimension of the VarCLR embedding~\cite{Chen2022}. The only exception to this is \CodexShort,
which we access indirectly through an HTTP API as explained in Section~\ref{sec:codex}.
In the \SinkPrediction and \BinaryClassification approaches, the hidden layer
dimensions are 500 and 250, respectively. We use the Adam optimizer with a
learning rate of 0.001. 
For \SinkPrediction, the training batch size is 256, and we stop the
training early when validation loss does not decrease in two epochs. We reserve
10\% of the training data for use as a validation set.
For \BinaryClassification, the training batch size is 32, and the
model stops when validation loss does not decrease in 50 epochs.
We use Radial Basis Function (RBF) as the kernel for the OC-SVM in the Novelty
Detection approach. The hyperparameters gamma and nu are 0.05 and 0.01,
respectively.
For the Codex language model, we set the temperature, frequency penalty, and
presence penalty to zero.

\subsection{RQ1: Identifying Unexpected Flows}
In this section, we explain the evaluation process for each approach and then
discuss the results for param-sink and logging flows. We compare the approaches
to each other, as well as to the two baselines (\Frequency and \CodeQL)
discussed above.

\subsubsection{Evaluation Process}

We investigate how well each approach can classify unexpected flows, using the
F1 score as our main metric. Table~\ref{table:method_summary} compares the
characteristics of the four neural approaches and the \Frequency baseline. All
approaches requiring a threshold are evaluated by plotting the precision-recall
curve (PR curve) showing the relationship between precision and recall for
different thresholds, and recording the best F-1 score obtainable.
We compute precision and recall w.r.t.\ source-sink pairs.

For \SinkPrediction, we train the model with all flows that have not been
labeled, using of these flows for 90\% training and 10\% for validation.
The labeled ground truth serves as the test set.
Since the model outputs probabilities, a threshold is required to make
a prediction. 

For \NoveltyDetection, we train one OC-SVM for each sink type (i.e., four for
param-sink flows, and one for logging flows) using the corresponding seed names.
We again use the ground truth as our test set, which may contain flows involving the seed names as these were chosen independently.
The model outputs a score for each flow, representing the distance between the
flow and the boundary of the OC-SVM, which again requires a threshold to derive
a prediction.

For \BinaryClassification, we also train one model per sink based on the
relevant flows from the balanced set. No threshold is needed since the model
directly classifies flows. We evaluate using 5-fold cross-validation with five different 80/20 splits, reporting
average precision, recall, and F1-score across the five folds for each model.

For \CodexShort, for each flow to evaluate, we randomly draw ten examples of the same sink from the balanced set to put in the prompt.
We ensure that the flow we ask it to evaluate is not among the ten examples embedded in the prompt.
As this model directly classifies whether the flow is expected, we again do not need a threshold and can directly compute the precision, recall, and F1-score based on the predictions.

Finally, \Frequency is similar to \SinkPrediction in that it is trained on all
param-sink flows and requires a threshold.
		
\subsubsection{Param-Sink Flows}
\label{s:QE}
For param-sink flows, all four approaches are evaluated on the random set
and the balanced set, with \Frequency as the baseline.
The summary results for the random set are shown in
Table~\ref{table:result_random}, with the best scores for each sink type
highlighted in bold.
We observe that \name{}'s approaches outperform the baseline for all sinks, with
\NoveltyDetection as the overall winner followed by \BinaryClassification. The
approaches perform especially well on command injection and reflected XSS with
F1 scores of 0.9--1.0, and around 0.8 for code injection. Effectiveness on path
traversal, on the other hand, is significantly worse with widely spread scores
peaking at 0.45. This is likely due to the random set having only eight examples
of unexpected path-traversal flow.

\begin{figure*}[t]
	\centering
	\includegraphics[width=\textwidth]{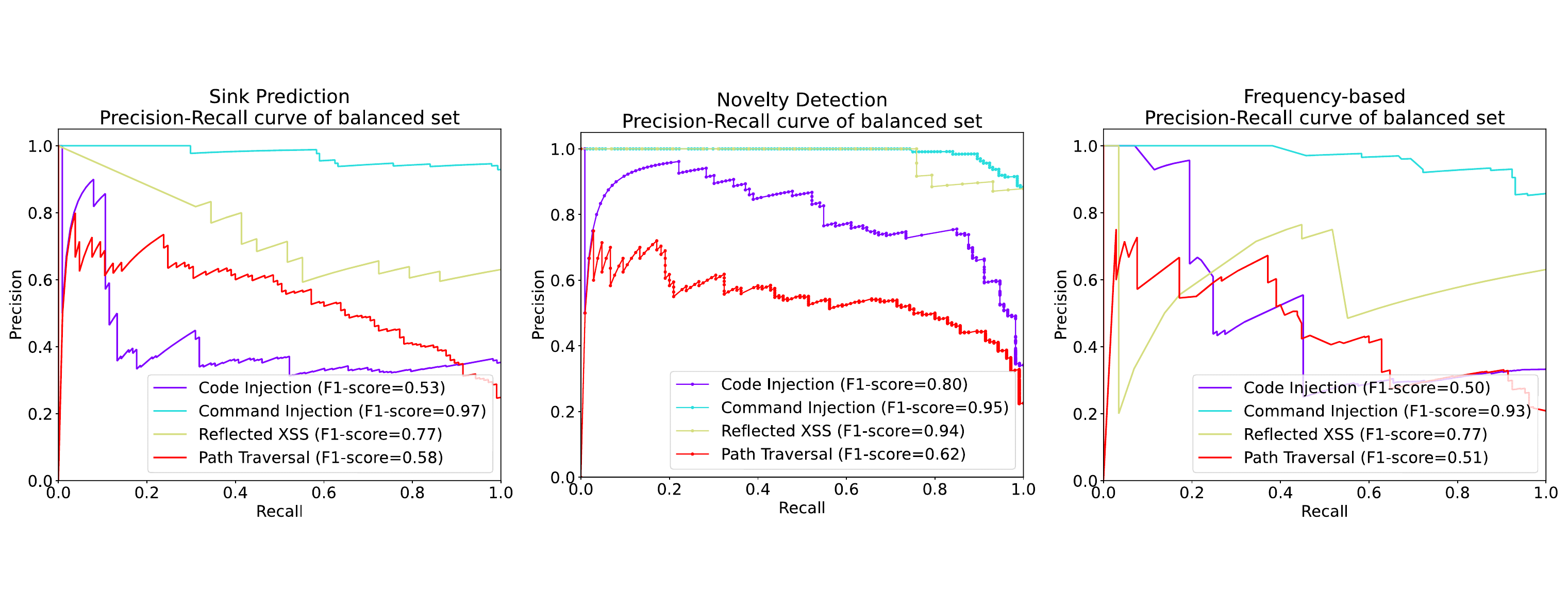}
	\vspace{-4em}
	\caption{PR curves on the balanced set for \SinkPrediction, \NoveltyDetection, and \Frequency (left to right).}
	\label{fig:balanced_set}
\end{figure*}

The result on the balanced set are shown in Table~\ref{table:result_whole}, with
PR curves in Figure~\ref{fig:balanced_set}. In this setup,
\BinaryClassification performs best, surpassing \NoveltyDetection by a
significant margin on code injection and path traversal. This could be due to
the balanced set containing more diverse names, giving \BinaryClassification an
opportunity to learn from the dataset, while \NoveltyDetection relies on a fixed
set of seed names. For the path traversal sink, all approaches perform better in
the balanced set than in the random set, which is likely due to having many more
(105) unexpected cases. \Frequency performs very well for command injection, but
this has to be interpreted with caution since the balanced set was partly
constructed from results of this very approach. Another interesting finding is
that \SinkPrediction performs much worse on code injection flows in the balanced
set than in the random set, as the F1-score drops from 0.82 to 0.53. This is due
to a decline in precision: while it is able to capture most of the unexpected
flows (108), it also includes many false positives (194). The rest of the
results are similar to the random set.

In summary, \name{} proves to be a significant improvement over the baseline
\Frequency approach. The two best approaches are \NoveltyDetection and \BinaryClassification, while \SinkPrediction does not work as well. Our interpretation
is that the observed likelihood of flow from a source to a sink does not help
understand how unexpected such a flow is. \CodexShort also does not perform as well
as the others, possibly because the randomly chosen samples embedded in the
prompt do not provide enough information.

Path traversal flows are the hardest to classify for all the approaches. This
seems to be because the names involved are very diverse, including file
operations (such as \texttt{move} and \texttt{rm}), file types, e.g., 
\texttt{png} and \texttt{mp4}), and other names that could represent certain
kinds of files (such as \texttt{log} and \texttt{config}). It is hard for the
models to learn all these concepts, reducing their effectiveness. The names for
other sink types are not as diverse, as shown, for example, by the excellent
score of the \NoveltyDetection model for command-injection flows in spite of
only having two seed names.

\begin{table}[]
	\centering
	\caption{Effectiveness of \name{} on the logging flows.}
	\setlength{\tabcolsep}{10pt}
	\label{table:result_logging}
	\begin{tabular}{@{}lrrr@{}}
		\toprule
		& \multicolumn{3}{c}{Logging Flows Dataset} \\ 
		\cmidrule(lr){2-4} Approach & \multicolumn{1}{l}{Precision} & \multicolumn{1}{l}{Recall} & \multicolumn{1}{l}{F1-Score} \\ 
		\midrule
		\SinkPrediction & 0.76 & 0.93 & 0.84 \\
		\NoveltyDetection & 0.81 & 0.93 & 0.87 \\
		\BinaryClassification & \textbf{0.90} & 0.94 & \textbf{0.92} \\
		\CodexShort & 0.78 & 0.96 & 0.86 \\
		\midrule
		\CodeQLShort & 0.79 & \textbf{0.97} & 0.87 \\
		\bottomrule
	\end{tabular}%
\end{table}

\begin{table}[ht]
	\centering
	\caption{Effectiveness of \name{} on the SecBench.js dataset.}
	\setlength{\tabcolsep}{4pt}
	\label{table:result_secbench}
	\begin{tabular}{@{}lrrr|rrr@{}}
		\toprule
		& \multicolumn{6}{c}{SecBench.js Dataset} \\ 
		\cmidrule{2-7} 
		& \multicolumn{3}{c|}{Code injection} & \multicolumn{3}{c}{Command injection} \\ 
		\cmidrule{2-7} 
		Approach & Precision & Recall & F1 & Precision & Recall & F1 \\ 
		\midrule
		\SinkPredictionShort & 0.97 & \textbf{0.88} & 0.92 & 1.00 & 0.90 & 0.95 \\
		\NoveltyDetectionShort & 0.96 & 0.75 & 0.84 & 0.99 & \textbf{0.98} & 0.98 \\
		\BinaryClassificationShort & 0.97 & 0.69 & 0.80 & 1.00 & 0.97 & 0.98 \\
		\CodexShort & 0.96 & 0.69 & 0.80 & 0.94 & 0.97 & 0.95 \\
		\midrule
		\FrequencyShort & 0.94 & 0.47 & 0.63 & 1.00 & 0.65 & 0.79 \\
		\bottomrule
	\end{tabular}%
\end{table}

\subsubsection{Logging Flows}

\begin{figure*}[t]
	\centering
	\includegraphics[width=0.9\textwidth]{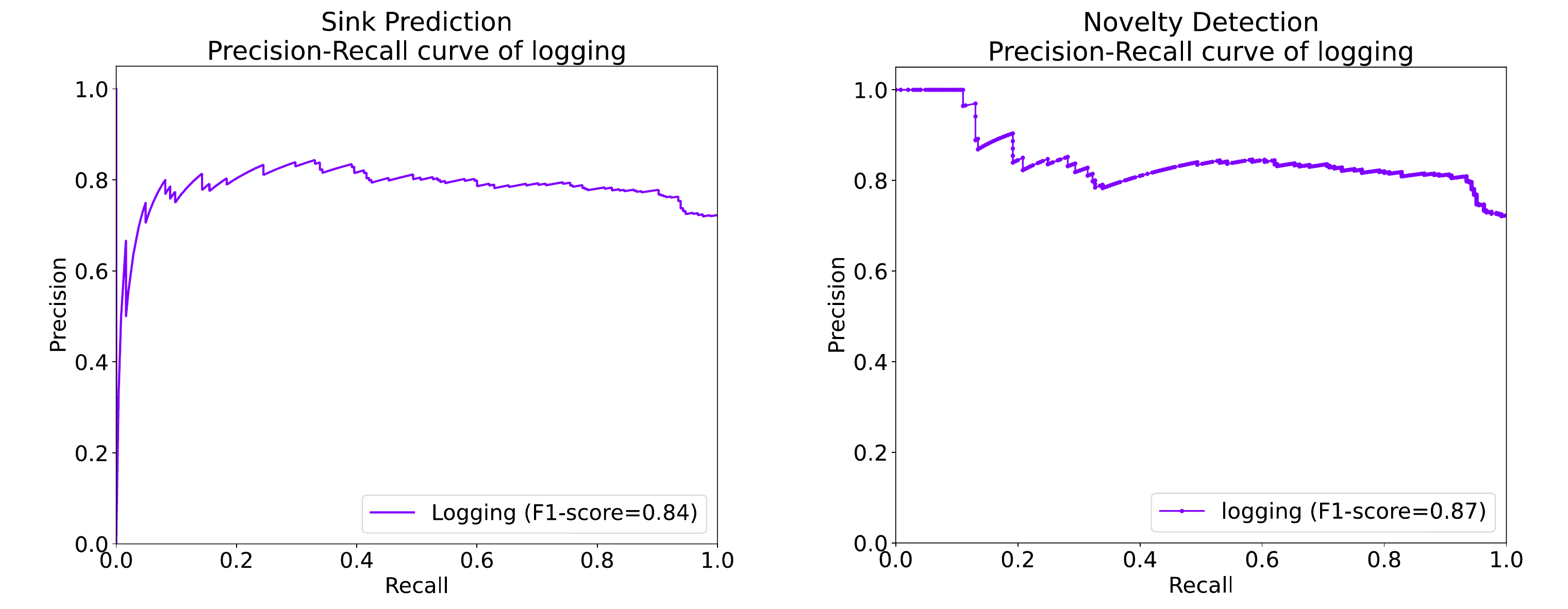}
	\caption{PR curves on the logging flows for \SinkPrediction and \NoveltyDetection (left to right).}
	\label{fig:logging}
\end{figure*}

The results of evaluating \name{} on the logging flows are shown in
Table~\ref{table:result_logging}, with PR curves in
Figure~\ref{fig:logging}. Overall, \BinaryClassification performs best, followed
by \NoveltyDetection and \CodeQL. The latter gives the highest recall, while the
latter have better precision. This suggests that regular expressions are fairly
effective, but they can be difficult to write and maintain while our
learning-based approach requires very little human intervention.

As a case in point, we discover that \name{} was able to correctly flag
clear-text logging of a variable named \code{passcode} as unexpected, while
CodeQL does not. This was surprising since the name matches one of the regular
expressions for finding sensitive data. However, it turns out that it
\emph{also} matches the regular expression used to filter out encrypted sources.
We report this to the CodeQL maintainers, who acknowledge it as a bug.

\subsection{RQ2: Real-World Vulnerabilities}
The results for RQ1 provide some evidence that \name{} is effective at spotting
unexpected flows, but this does not yet prove that it can be used to find
real-world vulnerabilities. To address this question, we evaluate \name{} on
the SecBench.js dataset, and also report some unexpected flows flagged in the
param-sink dataset to the maintainers of the corresponding projects to get their
input on whether the flows are indeed problematic.

\subsubsection{\name{} on Past Vulnerabilities} For evaluating on SecBench.js,
we use mostly the same evaluation methodology as shown Section~\ref{s:QE},
except for the threshold-based approaches where we use the threshold that gave
the best F-1 score on the balanced set in Section~\ref{s:QE}. This allows us to
assess if the threshold we set can generalize to a different dataset. We omit
the path-traversal sink type, since there is only one flow for this type in the
dataset.

We use recall as our main metric (Table~\ref{table:result_secbench}), since we
are mainly interested in whether \name{} would have been able to flag these
known vulnerabilities and since precision is already evaluated in RQ1. Somewhat surprisingly, \SinkPrediction has the best
recall for code injection. This turns out to be because the best threshold for
this approach on the balanced set is very high, favoring the SecBench.js
dataset where all but one code injection flows are unexpected. As before,
\NoveltyDetection does very well for command injection, closely followed by
\BinaryClassification.

\subsubsection{\name{} on Previously Unknown Vulnerabilities} In this section, we examine
how effective \name{} is at finding new vulnerabilities.
Inspecting the flows flagged by \name{} on the param-sink dataset, we find and
report a flow from a parameter called \code{name} of a function called
\code{locale} in the popular \code{moment} library to code-injection sink, as
mentioned in the introduction. The developers have acknowledged and fixed this
vulnerability, now known as CVE-2022-24785.
We furthermore create 16 pull requests on GitHub to fix other unexpected flows
that seemed undesirable but less critical. At the time of writing, seven of them
have been merged, eight are still open without feedback from developers, and one
pull request has been closed, with an alternative fix implemented.

Taken together, these results indicate that \name{} is indeed able to find
real-world vulnerabilities.

\begin{figure}[t]
	\centering
	\includegraphics[width=\columnwidth]{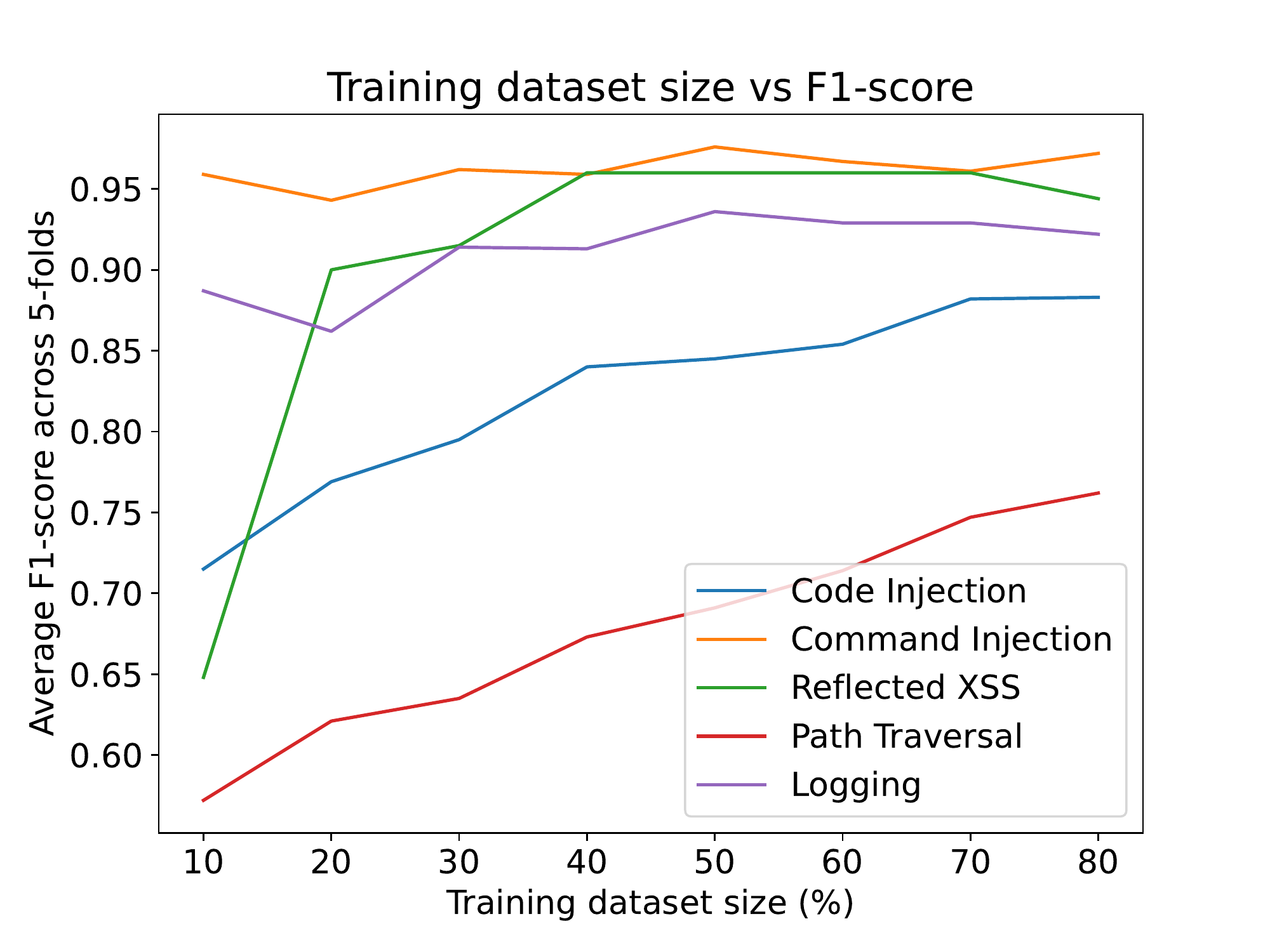}
	\vspace{-1em}
	\caption{Training set size versus model effectiveness.}
	\label{fig:train_size_exp}
\end{figure}

\subsection{RQ3: Human Efforts vs.\ Model Effectiveness}
From Table~\ref{table:result_random} and Table~\ref{table:result_whole}, we can
observe that \BinaryClassification is one of the approaches with the best
effectiveness. However, training the \BinaryClassification model requires a lot of
manual labeling, requiring human effort, which is a scarce resource. Therefore,
we would like to know how many labels are needed to achieve a good effectiveness
for \BinaryClassification.

In this experiment, we evaluate \BinaryClassification with different training
set sizes on the balanced set. Keeping the test set constant at 20\% of the
flows for a single sink type, we increase the training-set size from 10\% to
80\% in 10\% steps. 
The results in Figure~\ref{fig:train_size_exp} show that the F1 score generally
increases with training-set size, but for command injection even 10\% suffices.
\BinaryClassification with 10\% training data is, however, outperformed by
\NoveltyDetection with even less training data (viz, the seed names in
Table~\ref{tab:seed names}), showing that the former is a good choice if enough
labeled data is available, while the latter is more parsimonious.

\subsection{RQ4: Scalability}

One potential concern with our approach is the performance of the underlying
CodeQL query which has a very large set of sources. However, the query for the
param-sink flows dataset finished within LGTM.com's default timeout of four
hours on all but 73 projects ($>99.97\%$), and the one for the logging flows on
all but 49 ($>99.98\%$). This suggests that while \name{} might not scale to
truly massive projects, the CodeQL component is not usually a bottleneck.

For \BinaryClassification, our 5-fold cross-validation takes less than 5 minutes
to train and around 1.5 seconds to evaluate each model. The \SinkPrediction
model takes one and a half hour to train and around 20 seconds to evaluate. The
\NoveltyDetection model takes less than 3 seconds to train and less than 3
seconds to evaluate. For \CodexShort, the model is accessed via a rate-limited
REST API, providing one completion in 1.8 seconds, on average. This indicates
that, except for \CodexShort, the neural components of \name{} scale well, as
training takes at most a couple of hours, and the model can classify hundreds of
flows within seconds. 
We conduct our experiments on a server with 48 Intel Xeon CPU cores clocked at 2.2GHz, 250GB of RAM, and one NVIDIA Tesla V100 GPU. The \BinaryClassification and \SinkPrediction model use the GPU, whereas \NoveltyDetection and \CodexShort use the CPU.

\subsection{Threats to Validity}

While the results shown above are promising, there are several threats to
validity to keep in mind.
First, we only evaluate on five vulnerability types, and our results may not
generalize to other types. The corpus of code bases we consider is prescribed
by the selection of projects hosted on LGTM.com, potentially introducing bias.
Similarly, our evaluation is limited to JavaScript, and we cannot say for
certain that our approach would work for other languages.
Second, our evaluation relies on manual labeling, which is subject to human
error. We tried to mitigate this by involving independent labelers, but they
only labeled a small fraction of the data, leaving open the possibility of
bias in the remainder.
Third, our selection of seed names for \NoveltyDetection is based on
human judgment, again introducing a potential source of bias.
Finally, the balanced dataset contains more unexpected flows than we expect to actually occur in the wild, which may lead to overfitting.

\section{Related Work}
Our work fits in the general category of neural software
analysis~\cite{NeuralSoftwareAnalysis}, which is characterized by fuzziness of
available information and lack of well-defined correctness criteria, but also a
large amount of training data. Previous work in this area ranges tackles
problems as diverse as bug detection~\cite{oopsla2018-DeepBugs,Allamanis2018b,Vasic2019}, automated program repair~\cite{cacm2019-program-repair,Chen2019,Lutellier2020,Ye2022a,Li2020a},
code completion~\cite{Kim2021,Alon2019a,Chen2021},
probabilistic type inference~\cite{Hellendoorn2018,icse2019,fse2020,Allamanis2020}, call graph
pruning~\cite{Utture2022,Le-Cong2022}, emulating a dynamic taint
analysis~\cite{She2019}, and learning inference rules for static analyzers from
data~\cite{Bielik2017}. There has also been a lot of work on using
machine learning techniques to detect
vulnerabilities~\cite{Harer2018a,Li2018a,Li2021c,Fu2022}, though such approaches
are not yet precise enough in practice~\cite{Chakraborty2020}.

Unlike many of these works, however, our goal is not to produce a fully neural
end-to-end vulnerability detector. Instead, we enhance an existing
static analysis tool (CodeQL) with the help of machine learning techniques. In
this regard we stand in the tradition of systems like
Merlin~\cite{Livshits2009}, SuSi~\cite{susi}, Seldon~\cite{seldon}, and
USpec~\cite{Eberhardt2019}, which use machine learning to improve taint analyses
by identifying additional sources and sinks or inferring aliasing behavior.
However, those systems and our work have complementary goals.
While the above-mentioned approaches can be used to identify APIs that produce tainted data directly, \name{} finds parameters into which an unsuspecting user might pass tainted data, that is, the tainted data does not originate in the parameter, but somewhere else. Moreover, since we use natural-language information we can do this even if we have never seen an example of tainted data actually being passed to this parameter, thus allowing us to spot potential, not yet exploitable vulnerabilities.\footnote{While some features used by SuSi are based on natural language information, these are manually engineered and rely on very simple string matching, such as determining whether a method name starts with ``get''.}

We are, of course, not the first to realize and exploit the ``bi-modal'' structure of source code~\cite{Hindle2012,Allamanis2018,casalnuovo2020theory}.
For example, identifier names in particular have previously been used for bug
detection~\cite{oopsla2018-DeepBugs,icse2022-Nalin}, bug injection~\cite{fse2021}, type refinement~\cite{Dash2018}, and type prediction~\cite{icse2019}.  
However, we believe we are the first to systematically use natural language information for the purpose of improving taint analyses.

One could also view our approach as using neural techniques to identify false
positives, though in our case the underlying taint analysis is deliberately
imprecise and not expected to stand on its own. A recent contribution in this
direction is Kharkar et al.'s work~\cite{Kharkar2022}, which explores both a
feature-based and a neural approach to identifying false positives produced by
the Infer static analyzer~\cite{calcagno2015moving}. Their ideas relate to ours
in that their neural models implicitly use both code and natural language
information. Our work differs by specifically focusing on taint analysis and by
presenting other formulations of the learning task.

Another line of related work is on predicting heuristics used in static analyses.
For example, Jeong et al.\ learn heuristics for selecting methods likely to benefit from certain depth-levels of context-sensitive points-to analysis~\cite{Jeong2017}.
Others predict which loops to unroll without missing any bugs to report~\cite{Heo2017} and learn state-selection heuristics to reduce the cost of path-sensitivity~\cite{Ko2023icse}.
Such work ultimately aims at improving the efficiency of static analyses, whereas \name{}'s goal is to ensure that the warnings reported by the analysis are indeed relevant to developers.

The effectiveness of our approach depends crucially on the underlying word
embeddings. Common methods to train such embeddings include
FastText~\cite{Bojanowski2017}, word2vec~\cite{Mikolov2013} and
GloVe~\cite{glove}, but these are meant for natural language. We work with
variables names, which have substantially different
characteristics~\cite{icse2021}. Hence we use VarCLR~\cite{Chen2022}, which
specifically targets identifier names, and has been shown to outperform other
approaches in this domain.

\section{Conclusions}
This paper presents \name{}, a general framework for bimodal taint analysis
combining a static taint analysis identifying candidate flows with a neural
component identifying the problematic ones.
We instantiate this framework
on top of the CodeQL JavaScript analysis with four different neural components
(\BinaryClassification, \SinkPrediction, \NoveltyDetection, and \Codex), and
evaluate the resulting system on a large corpus of real-world JavaScript code
bases as well as a collection of known vulnerabilities.
Our findings show that
\name{} performs well in practice, identifying potentially problematic flows
with high accuracy and flagging known vulnerabilities with high recall.

For future work, we would like to investigate a tighter integration between the
static and the neural components, and also explore the broader applicability of
this technique to other languages.

\section*{Acknowledgments}
This work was supported by the European Research Council (ERC, grant agreement 851895), and by the German Research Foundation within the ConcSys and DeMoCo projects.

\bibliographystyle{ACM-Reference-Format}
\bibliography{referencesMP,referencesMore}


\begin{thebibliography}{54}


\ifx \showCODEN    \undefined \def \showCODEN     #1{\unskip}     \fi
\ifx \showDOI      \undefined \def \showDOI       #1{#1}\fi
\ifx \showISBNx    \undefined \def \showISBNx     #1{\unskip}     \fi
\ifx \showISBNxiii \undefined \def \showISBNxiii  #1{\unskip}     \fi
\ifx \showISSN     \undefined \def \showISSN      #1{\unskip}     \fi
\ifx \showLCCN     \undefined \def \showLCCN      #1{\unskip}     \fi
\ifx \shownote     \undefined \def \shownote      #1{#1}          \fi
\ifx \showarticletitle \undefined \def \showarticletitle #1{#1}   \fi
\ifx \showURL      \undefined \def \showURL       {\relax}        \fi
\providecommand\bibfield[2]{#2}
\providecommand\bibinfo[2]{#2}
\providecommand\natexlab[1]{#1}
\providecommand\showeprint[2][]{arXiv:#2}

\bibitem[Allamanis et~al\mbox{.}(2018a)]%
        {Allamanis2018}
\bibfield{author}{\bibinfo{person}{Miltiadis Allamanis},
  \bibinfo{person}{Earl~T Barr}, \bibinfo{person}{Premkumar Devanbu}, {and}
  \bibinfo{person}{Charles Sutton}.} \bibinfo{year}{2018}\natexlab{a}.
\newblock \showarticletitle{A survey of machine learning for big code and
  naturalness}.
\newblock \bibinfo{journal}{\emph{ACM Computing Surveys (CSUR)}}
  \bibinfo{volume}{51}, \bibinfo{number}{4} (\bibinfo{year}{2018}),
  \bibinfo{pages}{81}.
\newblock


\bibitem[Allamanis et~al\mbox{.}(2020)]%
        {Allamanis2020}
\bibfield{author}{\bibinfo{person}{Miltiadis Allamanis},
  \bibinfo{person}{Earl~T. Barr}, \bibinfo{person}{Soline Ducousso}, {and}
  \bibinfo{person}{Zheng Gao}.} \bibinfo{year}{2020}\natexlab{}.
\newblock \showarticletitle{Typilus: Neural Type Hints}. In
  \bibinfo{booktitle}{\emph{PLDI}}.
\newblock


\bibitem[Allamanis et~al\mbox{.}(2018b)]%
        {Allamanis2018b}
\bibfield{author}{\bibinfo{person}{Miltiadis Allamanis}, \bibinfo{person}{Marc
  Brockschmidt}, {and} \bibinfo{person}{Mahmoud Khademi}.}
  \bibinfo{year}{2018}\natexlab{b}.
\newblock \showarticletitle{Learning to Represent Programs with Graphs}. In
  \bibinfo{booktitle}{\emph{6th International Conference on Learning
  Representations, {ICLR} 2018, Vancouver, BC, Canada, April 30 - May 3, 2018,
  Conference Track Proceedings}}.
\newblock
\urldef\tempurl%
\url{https://openreview.net/forum?id=BJOFETxR-}
\showURL{%
\tempurl}


\bibitem[Alon et~al\mbox{.}(2019)]%
        {Alon2019a}
\bibfield{author}{\bibinfo{person}{Uri Alon}, \bibinfo{person}{Shaked Brody},
  \bibinfo{person}{Omer Levy}, {and} \bibinfo{person}{Eran Yahav}.}
  \bibinfo{year}{2019}\natexlab{}.
\newblock \showarticletitle{code2seq: Generating Sequences from Structured
  Representations of Code}. In \bibinfo{booktitle}{\emph{7th International
  Conference on Learning Representations, {ICLR} 2019, New Orleans, LA, USA,
  May 6-9, 2019}}.
\newblock
\urldef\tempurl%
\url{https://openreview.net/forum?id=H1gKYo09tX}
\showURL{%
\tempurl}


\bibitem[Avgustinov et~al\mbox{.}(2016)]%
        {ql}
\bibfield{author}{\bibinfo{person}{Pavel Avgustinov}, \bibinfo{person}{Oege de
  Moor}, \bibinfo{person}{Michael~Peyton Jones}, {and} \bibinfo{person}{Max
  Sch{\"{a}}fer}.} \bibinfo{year}{2016}\natexlab{}.
\newblock \showarticletitle{{QL: Object-oriented Queries on Relational Data}}.
  In \bibinfo{booktitle}{\emph{30th European Conference on Object-Oriented
  Programming, {ECOOP} 2016, July 18-22, 2016, Rome, Italy}}
  \emph{(\bibinfo{series}{LIPIcs}, Vol.~\bibinfo{volume}{56})},
  \bibfield{editor}{\bibinfo{person}{Shriram Krishnamurthi} {and}
  \bibinfo{person}{Benjamin~S. Lerner}} (Eds.). \bibinfo{publisher}{Schloss
  Dagstuhl - Leibniz-Zentrum f{\"{u}}r Informatik}, \bibinfo{pages}{2:1--2:25}.
\newblock
\urldef\tempurl%
\url{https://doi.org/10.4230/LIPIcs.ECOOP.2016.2}
\showDOI{\tempurl}


\bibitem[Barei{\ss} et~al\mbox{.}(2022)]%
        {codexStudy2022}
\bibfield{author}{\bibinfo{person}{Patrick Barei{\ss}},
  \bibinfo{person}{Beatriz Souza}, \bibinfo{person}{Marcelo d'Amorim}, {and}
  \bibinfo{person}{Michael Pradel}.} \bibinfo{year}{2022}\natexlab{}.
\newblock \showarticletitle{Code Generation Tools (Almost) for Free? {A} Study
  of Few-Shot, Pre-Trained Language Models on Code}.
\newblock \bibinfo{journal}{\emph{CoRR}}  \bibinfo{volume}{abs/2206.01335}
  (\bibinfo{year}{2022}).
\newblock
\urldef\tempurl%
\url{https://doi.org/10.48550/arXiv.2206.01335}
\showDOI{\tempurl}
\showeprint[arXiv]{2206.01335}


\bibitem[Bhuiyan et~al\mbox{.}(2022)]%
        {secbench}
\bibfield{author}{\bibinfo{person}{Masudul Bhuiyan}, \bibinfo{person}{Adithya
  Srinivas}, \bibinfo{person}{Nikos Vasilakis}, \bibinfo{person}{Michael
  Pradel}, {and} \bibinfo{person}{Cristian-Alexandru Staicu}.}
  \bibinfo{year}{2022}\natexlab{}.
\newblock \bibinfo{title}{{SecBench.js: An Executable Security Benchmark Suite
  for Server-Side JavaScript}}.
\newblock
\newblock
\urldef\tempurl%
\url{https://github.com/cristianstaicu/SecBench.js}
\showURL{%
\tempurl}


\bibitem[Bielik et~al\mbox{.}(2017)]%
        {Bielik2017}
\bibfield{author}{\bibinfo{person}{Pavol Bielik}, \bibinfo{person}{Veselin
  Raychev}, {and} \bibinfo{person}{Martin~T. Vechev}.}
  \bibinfo{year}{2017}\natexlab{}.
\newblock \showarticletitle{Learning a Static Analyzer from Data}. In
  \bibinfo{booktitle}{\emph{Computer Aided Verification - 29th International
  Conference, {CAV} 2017, Heidelberg, Germany, July 24-28, 2017, Proceedings,
  Part {I}}} \emph{(\bibinfo{series}{Lecture Notes in Computer Science},
  Vol.~\bibinfo{volume}{10426})}, \bibfield{editor}{\bibinfo{person}{Rupak
  Majumdar} {and} \bibinfo{person}{Viktor Kuncak}} (Eds.).
  \bibinfo{publisher}{Springer}, \bibinfo{pages}{233--253}.
\newblock
\urldef\tempurl%
\url{https://doi.org/10.1007/978-3-319-63387-9\_12}
\showDOI{\tempurl}


\bibitem[Bojanowski et~al\mbox{.}(2017)]%
        {Bojanowski2017}
\bibfield{author}{\bibinfo{person}{Piotr Bojanowski}, \bibinfo{person}{Edouard
  Grave}, \bibinfo{person}{Armand Joulin}, {and} \bibinfo{person}{Tomas
  Mikolov}.} \bibinfo{year}{2017}\natexlab{}.
\newblock \showarticletitle{Enriching Word Vectors with Subword Information}.
\newblock \bibinfo{journal}{\emph{{TACL}}}  \bibinfo{volume}{5}
  (\bibinfo{year}{2017}), \bibinfo{pages}{135--146}.
\newblock
\urldef\tempurl%
\url{https://transacl.org/ojs/index.php/tacl/article/view/999}
\showURL{%
\tempurl}


\bibitem[Brown et~al\mbox{.}(2020)]%
        {Brown2020}
\bibfield{author}{\bibinfo{person}{Tom~B. Brown}, \bibinfo{person}{Benjamin
  Mann}, \bibinfo{person}{Nick Ryder}, \bibinfo{person}{Melanie Subbiah},
  \bibinfo{person}{Jared Kaplan}, \bibinfo{person}{Prafulla Dhariwal},
  \bibinfo{person}{Arvind Neelakantan}, \bibinfo{person}{Pranav Shyam},
  \bibinfo{person}{Girish Sastry}, \bibinfo{person}{Amanda Askell},
  \bibinfo{person}{Sandhini Agarwal}, \bibinfo{person}{Ariel Herbert{-}Voss},
  \bibinfo{person}{Gretchen Krueger}, \bibinfo{person}{Tom Henighan},
  \bibinfo{person}{Rewon Child}, \bibinfo{person}{Aditya Ramesh},
  \bibinfo{person}{Daniel~M. Ziegler}, \bibinfo{person}{Jeffrey Wu},
  \bibinfo{person}{Clemens Winter}, \bibinfo{person}{Christopher Hesse},
  \bibinfo{person}{Mark Chen}, \bibinfo{person}{Eric Sigler},
  \bibinfo{person}{Mateusz Litwin}, \bibinfo{person}{Scott Gray},
  \bibinfo{person}{Benjamin Chess}, \bibinfo{person}{Jack Clark},
  \bibinfo{person}{Christopher Berner}, \bibinfo{person}{Sam McCandlish},
  \bibinfo{person}{Alec Radford}, \bibinfo{person}{Ilya Sutskever}, {and}
  \bibinfo{person}{Dario Amodei}.} \bibinfo{year}{2020}\natexlab{}.
\newblock \showarticletitle{Language Models are Few-Shot Learners}. In
  \bibinfo{booktitle}{\emph{Advances in Neural Information Processing Systems
  33: Annual Conference on Neural Information Processing Systems 2020, NeurIPS
  2020, December 6-12, 2020, virtual}}, \bibfield{editor}{\bibinfo{person}{Hugo
  Larochelle}, \bibinfo{person}{Marc'Aurelio Ranzato}, \bibinfo{person}{Raia
  Hadsell}, \bibinfo{person}{Maria{-}Florina Balcan}, {and}
  \bibinfo{person}{Hsuan{-}Tien Lin}} (Eds.).
\newblock
\urldef\tempurl%
\url{https://proceedings.neurips.cc/paper/2020/hash/1457c0d6bfcb4967418bfb8ac142f64a-Abstract.html}
\showURL{%
\tempurl}


\bibitem[Calcagno et~al\mbox{.}(2015)]%
        {calcagno2015moving}
\bibfield{author}{\bibinfo{person}{Cristiano Calcagno}, \bibinfo{person}{Dino
  Distefano}, \bibinfo{person}{J{\'e}r{\'e}my Dubreil},
  \bibinfo{person}{Dominik Gabi}, \bibinfo{person}{Pieter Hooimeijer},
  \bibinfo{person}{Martino Luca}, \bibinfo{person}{Peter O’Hearn},
  \bibinfo{person}{Irene Papakonstantinou}, \bibinfo{person}{Jim Purbrick},
  {and} \bibinfo{person}{Dulma Rodriguez}.} \bibinfo{year}{2015}\natexlab{}.
\newblock \showarticletitle{Moving fast with software verification}. In
  \bibinfo{booktitle}{\emph{NASA Formal Methods Symposium}}. Springer,
  \bibinfo{pages}{3--11}.
\newblock


\bibitem[Casalnuovo et~al\mbox{.}(2020)]%
        {casalnuovo2020theory}
\bibfield{author}{\bibinfo{person}{Casey Casalnuovo}, \bibinfo{person}{Earl~T
  Barr}, \bibinfo{person}{Santanu~Kumar Dash}, \bibinfo{person}{Prem Devanbu},
  {and} \bibinfo{person}{Emily Morgan}.} \bibinfo{year}{2020}\natexlab{}.
\newblock \showarticletitle{A theory of dual channel constraints}. In
  \bibinfo{booktitle}{\emph{2020 IEEE/ACM 42nd International Conference on
  Software Engineering: New Ideas and Emerging Results (ICSE-NIER)}}. IEEE,
  \bibinfo{pages}{25--28}.
\newblock


\bibitem[Chakraborty et~al\mbox{.}(2020)]%
        {Chakraborty2020}
\bibfield{author}{\bibinfo{person}{Saikat Chakraborty}, \bibinfo{person}{Rahul
  Krishna}, \bibinfo{person}{Yangruibo Ding}, {and} \bibinfo{person}{Baishakhi
  Ray}.} \bibinfo{year}{2020}\natexlab{}.
\newblock \showarticletitle{Deep Learning based Vulnerability Detection: Are We
  There Yet?}
\newblock \bibinfo{journal}{\emph{CoRR}}  \bibinfo{volume}{abs/2009.07235}
  (\bibinfo{year}{2020}).
\newblock
\showeprint[arXiv]{2009.07235}
\urldef\tempurl%
\url{https://arxiv.org/abs/2009.07235}
\showURL{%
\tempurl}


\bibitem[Chen et~al\mbox{.}(2021)]%
        {Chen2021}
\bibfield{author}{\bibinfo{person}{Mark Chen}, \bibinfo{person}{Jerry Tworek},
  \bibinfo{person}{Heewoo Jun}, \bibinfo{person}{Qiming Yuan},
  \bibinfo{person}{Henrique~Ponde de Oliveira~Pinto}, \bibinfo{person}{Jared
  Kaplan}, \bibinfo{person}{Harrison Edwards}, \bibinfo{person}{Yuri Burda},
  \bibinfo{person}{Nicholas Joseph}, \bibinfo{person}{Greg Brockman},
  \bibinfo{person}{Alex Ray}, \bibinfo{person}{Raul Puri},
  \bibinfo{person}{Gretchen Krueger}, \bibinfo{person}{Michael Petrov},
  \bibinfo{person}{Heidy Khlaaf}, \bibinfo{person}{Girish Sastry},
  \bibinfo{person}{Pamela Mishkin}, \bibinfo{person}{Brooke Chan},
  \bibinfo{person}{Scott Gray}, \bibinfo{person}{Nick Ryder},
  \bibinfo{person}{Mikhail Pavlov}, \bibinfo{person}{Alethea Power},
  \bibinfo{person}{Lukasz Kaiser}, \bibinfo{person}{Mohammad Bavarian},
  \bibinfo{person}{Clemens Winter}, \bibinfo{person}{Philippe Tillet},
  \bibinfo{person}{Felipe~Petroski Such}, \bibinfo{person}{Dave Cummings},
  \bibinfo{person}{Matthias Plappert}, \bibinfo{person}{Fotios Chantzis},
  \bibinfo{person}{Elizabeth Barnes}, \bibinfo{person}{Ariel Herbert{-}Voss},
  \bibinfo{person}{William~Hebgen Guss}, \bibinfo{person}{Alex Nichol},
  \bibinfo{person}{Alex Paino}, \bibinfo{person}{Nikolas Tezak},
  \bibinfo{person}{Jie Tang}, \bibinfo{person}{Igor Babuschkin},
  \bibinfo{person}{Suchir Balaji}, \bibinfo{person}{Shantanu Jain},
  \bibinfo{person}{William Saunders}, \bibinfo{person}{Christopher Hesse},
  \bibinfo{person}{Andrew~N. Carr}, \bibinfo{person}{Jan Leike},
  \bibinfo{person}{Joshua Achiam}, \bibinfo{person}{Vedant Misra},
  \bibinfo{person}{Evan Morikawa}, \bibinfo{person}{Alec Radford},
  \bibinfo{person}{Matthew Knight}, \bibinfo{person}{Miles Brundage},
  \bibinfo{person}{Mira Murati}, \bibinfo{person}{Katie Mayer},
  \bibinfo{person}{Peter Welinder}, \bibinfo{person}{Bob McGrew},
  \bibinfo{person}{Dario Amodei}, \bibinfo{person}{Sam McCandlish},
  \bibinfo{person}{Ilya Sutskever}, {and} \bibinfo{person}{Wojciech Zaremba}.}
  \bibinfo{year}{2021}\natexlab{}.
\newblock \showarticletitle{Evaluating Large Language Models Trained on Code}.
\newblock \bibinfo{journal}{\emph{CoRR}}  \bibinfo{volume}{abs/2107.03374}
  (\bibinfo{year}{2021}).
\newblock
\showeprint[arXiv]{2107.03374}
\urldef\tempurl%
\url{https://arxiv.org/abs/2107.03374}
\showURL{%
\tempurl}


\bibitem[Chen et~al\mbox{.}(2022)]%
        {Chen2022}
\bibfield{author}{\bibinfo{person}{Qibin Chen}, \bibinfo{person}{Jeremy
  Lacomis}, \bibinfo{person}{Edward~J. Schwartz}, \bibinfo{person}{Graham
  Neubig}, \bibinfo{person}{Bogdan Vasilescu}, {and} \bibinfo{person}{Claire~Le
  Goues}.} \bibinfo{year}{2022}\natexlab{}.
\newblock \showarticletitle{VarCLR: Variable Semantic Representation
  Pre-training via Contrastive Learning}. In \bibinfo{booktitle}{\emph{ICSE}}.
\newblock


\bibitem[Chen et~al\mbox{.}(2019)]%
        {Chen2019}
\bibfield{author}{\bibinfo{person}{Zimin Chen}, \bibinfo{person}{Steve
  Kommrusch}, \bibinfo{person}{Michele Tufano},
  \bibinfo{person}{Louis{-}No{\"{e}}l Pouchet}, \bibinfo{person}{Denys
  Poshyvanyk}, {and} \bibinfo{person}{Martin Monperrus}.}
  \bibinfo{year}{2019}\natexlab{}.
\newblock \showarticletitle{SequenceR: Sequence-to-Sequence Learning for
  End-to-End Program Repair}.
\newblock \bibinfo{journal}{\emph{IEEE TSE}} (\bibinfo{year}{2019}).
\newblock


\bibitem[Chibotaru et~al\mbox{.}(2019)]%
        {seldon}
\bibfield{author}{\bibinfo{person}{Victor Chibotaru}, \bibinfo{person}{Benjamin
  Bichsel}, \bibinfo{person}{Veselin Raychev}, {and} \bibinfo{person}{Martin~T.
  Vechev}.} \bibinfo{year}{2019}\natexlab{}.
\newblock \showarticletitle{{Scalable Taint Specification Inference with Big
  Code}}. In \bibinfo{booktitle}{\emph{Proceedings of the 40th {ACM} {SIGPLAN}
  Conference on Programming Language Design and Implementation, {PLDI} 2019,
  Phoenix, AZ, USA, June 22-26, 2019}},
  \bibfield{editor}{\bibinfo{person}{Kathryn~S. McKinley} {and}
  \bibinfo{person}{Kathleen Fisher}} (Eds.). \bibinfo{publisher}{{ACM}},
  \bibinfo{pages}{760--774}.
\newblock
\urldef\tempurl%
\url{https://doi.org/10.1145/3314221.3314648}
\showDOI{\tempurl}


\bibitem[Dash et~al\mbox{.}(2018)]%
        {Dash2018}
\bibfield{author}{\bibinfo{person}{Santanu~Kumar Dash},
  \bibinfo{person}{Miltiadis Allamanis}, {and} \bibinfo{person}{Earl~T. Barr}.}
  \bibinfo{year}{2018}\natexlab{}.
\newblock \showarticletitle{RefiNym: Using Names to Refine Types}. In
  \bibinfo{booktitle}{\emph{ESEC/FSE}}.
\newblock


\bibitem[Eberhardt et~al\mbox{.}(2019)]%
        {Eberhardt2019}
\bibfield{author}{\bibinfo{person}{Jan Eberhardt}, \bibinfo{person}{Samuel
  Steffen}, \bibinfo{person}{Veselin Raychev}, {and} \bibinfo{person}{Martin~T.
  Vechev}.} \bibinfo{year}{2019}\natexlab{}.
\newblock \showarticletitle{Unsupervised learning of {API} aliasing
  specifications}. In \bibinfo{booktitle}{\emph{Proceedings of the 40th {ACM}
  {SIGPLAN} Conference on Programming Language Design and Implementation,
  {PLDI} 2019, Phoenix, AZ, USA, June 22-26, 2019.}} \bibinfo{pages}{745--759}.
\newblock
\urldef\tempurl%
\url{https://doi.org/10.1145/3314221.3314640}
\showDOI{\tempurl}


\bibitem[Fu and Tantithamthavorn(2022)]%
        {Fu2022}
\bibfield{author}{\bibinfo{person}{Michael Fu} {and} \bibinfo{person}{Chakkrit
  Tantithamthavorn}.} \bibinfo{year}{2022}\natexlab{}.
\newblock \showarticletitle{LineVul: A Transformer-based Line-Level
  Vulnerability Prediction}. In \bibinfo{booktitle}{\emph{MSR}}.
\newblock


\bibitem[Gazit(2022)]%
        {atm}
\bibfield{author}{\bibinfo{person}{Tiferet Gazit}.}
  \bibinfo{year}{2022}\natexlab{}.
\newblock \bibinfo{title}{{Leveraging machine learning to find security
  vulnerabilities}}.
\newblock
\newblock
\urldef\tempurl%
\url{https://github.blog/2022-02-17-leveraging-machine-learning-find-security-vulnerabilities/}
\showURL{%
\tempurl}


\bibitem[GitHub(2022)]%
        {codeql}
\bibfield{author}{\bibinfo{person}{GitHub}.} \bibinfo{year}{2022}\natexlab{}.
\newblock \bibinfo{title}{{CodeQL}}.
\newblock
\newblock
\urldef\tempurl%
\url{https://codeql.github.com/}
\showURL{%
\tempurl}


\bibitem[Harer et~al\mbox{.}(2018)]%
        {Harer2018a}
\bibfield{author}{\bibinfo{person}{Jacob~A. Harer}, \bibinfo{person}{Louis~Y.
  Kim}, \bibinfo{person}{Rebecca~L. Russell}, \bibinfo{person}{Onur Ozdemir},
  \bibinfo{person}{Leonard~R. Kosta}, \bibinfo{person}{Akshay Rangamani},
  \bibinfo{person}{Lei~H. Hamilton}, \bibinfo{person}{Gabriel~I. Centeno},
  \bibinfo{person}{Jonathan~R. Key}, \bibinfo{person}{Paul~M. Ellingwood},
  \bibinfo{person}{Marc~W. McConley}, \bibinfo{person}{Jeffrey~M. Opper},
  \bibinfo{person}{Sang~Peter Chin}, {and} \bibinfo{person}{Tomo Lazovich}.}
  \bibinfo{year}{2018}\natexlab{}.
\newblock \showarticletitle{Automated software vulnerability detection with
  machine learning}.
\newblock \bibinfo{journal}{\emph{CoRR}}  \bibinfo{volume}{abs/1803.04497}
  (\bibinfo{year}{2018}).
\newblock
\showeprint[arxiv]{1803.04497}
\urldef\tempurl%
\url{http://arxiv.org/abs/1803.04497}
\showURL{%
\tempurl}


\bibitem[Hellendoorn et~al\mbox{.}(2018)]%
        {Hellendoorn2018}
\bibfield{author}{\bibinfo{person}{Vincent~J. Hellendoorn},
  \bibinfo{person}{Christian Bird}, \bibinfo{person}{Earl~T. Barr}, {and}
  \bibinfo{person}{Miltiadis Allamanis}.} \bibinfo{year}{2018}\natexlab{}.
\newblock \showarticletitle{Deep learning type inference}. In
  \bibinfo{booktitle}{\emph{Proceedings of the 2018 {ACM} Joint Meeting on
  European Software Engineering Conference and Symposium on the Foundations of
  Software Engineering, {ESEC/SIGSOFT} {FSE} 2018, Lake Buena Vista, FL, USA,
  November 04-09, 2018}}, \bibfield{editor}{\bibinfo{person}{Gary~T. Leavens},
  \bibinfo{person}{Alessandro Garcia}, {and} \bibinfo{person}{Corina~S.
  Pasareanu}} (Eds.). \bibinfo{publisher}{{ACM}}, \bibinfo{pages}{152--162}.
\newblock
\urldef\tempurl%
\url{https://doi.org/10.1145/3236024.3236051}
\showDOI{\tempurl}


\bibitem[Heo et~al\mbox{.}(2017)]%
        {Heo2017}
\bibfield{author}{\bibinfo{person}{Kihong Heo}, \bibinfo{person}{Hakjoo Oh},
  {and} \bibinfo{person}{Kwangkeun Yi}.} \bibinfo{year}{2017}\natexlab{}.
\newblock \showarticletitle{Machine-learning-guided selectively unsound static
  analysis}. In \bibinfo{booktitle}{\emph{Proceedings of the 39th International
  Conference on Software Engineering, {ICSE} 2017, Buenos Aires, Argentina, May
  20-28, 2017}}, \bibfield{editor}{\bibinfo{person}{Sebasti{\'{a}}n Uchitel},
  \bibinfo{person}{Alessandro Orso}, {and} \bibinfo{person}{Martin~P.
  Robillard}} (Eds.). \bibinfo{publisher}{{IEEE} / {ACM}},
  \bibinfo{pages}{519--529}.
\newblock
\urldef\tempurl%
\url{https://doi.org/10.1109/ICSE.2017.54}
\showDOI{\tempurl}


\bibitem[Hindle et~al\mbox{.}(2012)]%
        {Hindle2012}
\bibfield{author}{\bibinfo{person}{Abram Hindle}, \bibinfo{person}{Earl~T.
  Barr}, \bibinfo{person}{Zhendong Su}, \bibinfo{person}{Mark Gabel}, {and}
  \bibinfo{person}{Premkumar~T. Devanbu}.} \bibinfo{year}{2012}\natexlab{}.
\newblock \showarticletitle{On the naturalness of software}. In
  \bibinfo{booktitle}{\emph{34th International Conference on Software
  Engineering, {ICSE} 2012, June 2-9, 2012, Zurich, Switzerland}}.
  \bibinfo{pages}{837--847}.
\newblock


\bibitem[Jain et~al\mbox{.}(2022)]%
        {Jain2022}
\bibfield{author}{\bibinfo{person}{Naman Jain}, \bibinfo{person}{Skanda
  Vaidyanath}, \bibinfo{person}{Arun Iyer}, \bibinfo{person}{Nagarajan
  Natarajan}, \bibinfo{person}{Suresh Parthasarathy}, \bibinfo{person}{Sriram
  Rajamani}, {and} \bibinfo{person}{Rahul Sharma}.}
  \bibinfo{year}{2022}\natexlab{}.
\newblock \showarticletitle{Jigsaw: Large Language Models meet Program
  Synthesis}. In \bibinfo{booktitle}{\emph{ICSE}}.
\newblock


\bibitem[Jeong et~al\mbox{.}(2017)]%
        {Jeong2017}
\bibfield{author}{\bibinfo{person}{Sehun Jeong}, \bibinfo{person}{Minseok
  Jeon}, \bibinfo{person}{Sung~Deok Cha}, {and} \bibinfo{person}{Hakjoo Oh}.}
  \bibinfo{year}{2017}\natexlab{}.
\newblock \showarticletitle{Data-driven context-sensitivity for points-to
  analysis}.
\newblock \bibinfo{journal}{\emph{Proc. {ACM} Program. Lang.}}
  \bibinfo{volume}{1}, \bibinfo{number}{{OOPSLA}} (\bibinfo{year}{2017}),
  \bibinfo{pages}{100:1--100:28}.
\newblock
\urldef\tempurl%
\url{https://doi.org/10.1145/3133924}
\showDOI{\tempurl}


\bibitem[Joshi et~al\mbox{.}(2022)]%
        {Joshi2022}
\bibfield{author}{\bibinfo{person}{Harshit Joshi}, \bibinfo{person}{José
  Cambronero}, \bibinfo{person}{Sumit Gulwani}, \bibinfo{person}{Vu Le},
  \bibinfo{person}{Ivan Radicek}, {and} \bibinfo{person}{Gust Verbruggen}.}
  \bibinfo{year}{2022}\natexlab{}.
\newblock \bibinfo{title}{Repair Is Nearly Generation: Multilingual Program
  Repair with LLMs}.
\newblock
\newblock
\urldef\tempurl%
\url{https://doi.org/10.48550/ARXIV.2208.11640}
\showDOI{\tempurl}


\bibitem[Kharkar et~al\mbox{.}(2022)]%
        {Kharkar2022}
\bibfield{author}{\bibinfo{person}{Anant Kharkar},
  \bibinfo{person}{Roshanak~Zilouchian Moghaddam}, \bibinfo{person}{Matthew
  Jin}, \bibinfo{person}{Xiaoyu Liu}, \bibinfo{person}{Xin Shi},
  \bibinfo{person}{Colin Clement}, {and} \bibinfo{person}{Neel Sundaresan}.}
  \bibinfo{year}{2022}\natexlab{}.
\newblock \showarticletitle{Learning to Reduce False Positives in Analytic Bug
  Detectors}. In \bibinfo{booktitle}{\emph{ICSE}}.
\newblock


\bibitem[Kim et~al\mbox{.}(2021)]%
        {Kim2021}
\bibfield{author}{\bibinfo{person}{Seohyun Kim}, \bibinfo{person}{Jinman Zhao},
  \bibinfo{person}{Yuchi Tian}, {and} \bibinfo{person}{Satish Chandra}.}
  \bibinfo{year}{2021}\natexlab{}.
\newblock \showarticletitle{Code Prediction by Feeding Trees to Transformers}.
  In \bibinfo{booktitle}{\emph{ICSE}}.
\newblock


\bibitem[Ko and Oh(2023)]%
        {Ko2023icse}
\bibfield{author}{\bibinfo{person}{Yoonseok Ko} {and} \bibinfo{person}{Hakjoo
  Oh}.} \bibinfo{year}{2023}\natexlab{}.
\newblock \showarticletitle{Learning to Boost Disjunctive Static Bug-Finders}.
  In \bibinfo{booktitle}{\emph{ICSE}}.
\newblock


\bibitem[Le-Cong et~al\mbox{.}(2022)]%
        {Le-Cong2022}
\bibfield{author}{\bibinfo{person}{Thanh Le-Cong}, \bibinfo{person}{Hong~Jin
  Kang}, \bibinfo{person}{Truong~Giang Nguyen}, \bibinfo{person}{Stefanus~Agus
  Haryono}, \bibinfo{person}{David Lo}, \bibinfo{person}{Xuan-Bach~D. Le},
  {and} \bibinfo{person}{Quyet~Thang Huynh}.} \bibinfo{year}{2022}\natexlab{}.
\newblock \showarticletitle{AutoPruner: Transformer-Based Call Graph Pruning}.
  In \bibinfo{booktitle}{\emph{ESEC/FSE}}.
\newblock


\bibitem[{Le Goues} et~al\mbox{.}(2019)]%
        {cacm2019-program-repair}
\bibfield{author}{\bibinfo{person}{Claire {Le Goues}}, \bibinfo{person}{Michael
  Pradel}, {and} \bibinfo{person}{Abhik Roychoudhury}.}
  \bibinfo{year}{2019}\natexlab{}.
\newblock \showarticletitle{Automated program repair}.
\newblock \bibinfo{journal}{\emph{Commun. {ACM}}} \bibinfo{volume}{62},
  \bibinfo{number}{12} (\bibinfo{year}{2019}), \bibinfo{pages}{56--65}.
\newblock
\urldef\tempurl%
\url{https://doi.org/10.1145/3318162}
\showDOI{\tempurl}


\bibitem[Li et~al\mbox{.}(2020)]%
        {Li2020a}
\bibfield{author}{\bibinfo{person}{Yi Li}, \bibinfo{person}{Shaohua Wang},
  {and} \bibinfo{person}{Tien~N. Nguyen}.} \bibinfo{year}{2020}\natexlab{}.
\newblock \showarticletitle{DLFix: Context-based Code Transformation Learning
  for Automated Program Repair}. In \bibinfo{booktitle}{\emph{ICSE}}.
\newblock


\bibitem[Li et~al\mbox{.}(2021)]%
        {Li2021c}
\bibfield{author}{\bibinfo{person}{Yi Li}, \bibinfo{person}{Shaohua Wang},
  {and} \bibinfo{person}{Tien~N. Nguyen}.} \bibinfo{year}{2021}\natexlab{}.
\newblock \showarticletitle{Vulnerability detection with fine-grained
  interpretations}. In \bibinfo{booktitle}{\emph{{ESEC/FSE} '21: 29th {ACM}
  Joint European Software Engineering Conference and Symposium on the
  Foundations of Software Engineering, Athens, Greece, August 23-28, 2021}},
  \bibfield{editor}{\bibinfo{person}{Diomidis Spinellis},
  \bibinfo{person}{Georgios Gousios}, \bibinfo{person}{Marsha Chechik}, {and}
  \bibinfo{person}{Massimiliano~Di Penta}} (Eds.). \bibinfo{publisher}{{ACM}},
  \bibinfo{pages}{292--303}.
\newblock
\urldef\tempurl%
\url{https://doi.org/10.1145/3468264.3468597}
\showDOI{\tempurl}


\bibitem[Li et~al\mbox{.}(2018)]%
        {Li2018a}
\bibfield{author}{\bibinfo{person}{Zhen Li}, \bibinfo{person}{Shouhuai~Xu
  Deqing Zou~and}, \bibinfo{person}{Xinyu Ou}, \bibinfo{person}{Hai Jin},
  \bibinfo{person}{Sujuan Wang}, \bibinfo{person}{Zhijun Deng}, {and}
  \bibinfo{person}{Yuyi Zhong}.} \bibinfo{year}{2018}\natexlab{}.
\newblock \showarticletitle{{VulDeePecker}: A Deep Learning-Based System for
  Vulnerability Detection}. In \bibinfo{booktitle}{\emph{NDSS}}.
\newblock


\bibitem[Livshits et~al\mbox{.}(2009)]%
        {Livshits2009}
\bibfield{author}{\bibinfo{person}{V.~Benjamin Livshits},
  \bibinfo{person}{Aditya~V. Nori}, \bibinfo{person}{Sriram~K. Rajamani}, {and}
  \bibinfo{person}{Anindya Banerjee}.} \bibinfo{year}{2009}\natexlab{}.
\newblock \showarticletitle{Merlin: specification inference for explicit
  information flow problems}. In \bibinfo{booktitle}{\emph{Proceedings of the
  2009 {ACM} {SIGPLAN} Conference on Programming Language Design and
  Implementation, {PLDI} 2009, Dublin, Ireland, June 15-21, 2009}}.
  \bibinfo{pages}{75--86}.
\newblock


\bibitem[Lutellier et~al\mbox{.}(2020)]%
        {Lutellier2020}
\bibfield{author}{\bibinfo{person}{Thibaud Lutellier},
  \bibinfo{person}{Hung~Viet Pham}, \bibinfo{person}{Lawrence Pang},
  \bibinfo{person}{Yitong Li}, \bibinfo{person}{Moshi Wei}, {and}
  \bibinfo{person}{Lin Tan}.} \bibinfo{year}{2020}\natexlab{}.
\newblock \showarticletitle{CoCoNuT: combining context-aware neural translation
  models using ensemble for program repair}. In
  \bibinfo{booktitle}{\emph{{ISSTA} '20: 29th {ACM} {SIGSOFT} International
  Symposium on Software Testing and Analysis, Virtual Event, USA, July 18-22,
  2020}}, \bibfield{editor}{\bibinfo{person}{Sarfraz Khurshid} {and}
  \bibinfo{person}{Corina~S. Pasareanu}} (Eds.). \bibinfo{publisher}{{ACM}},
  \bibinfo{pages}{101--114}.
\newblock
\urldef\tempurl%
\url{https://doi.org/10.1145/3395363.3397369}
\showDOI{\tempurl}


\bibitem[Malik et~al\mbox{.}(2019)]%
        {icse2019}
\bibfield{author}{\bibinfo{person}{Rabee~Sohail Malik}, \bibinfo{person}{Jibesh
  Patra}, {and} \bibinfo{person}{Michael Pradel}.}
  \bibinfo{year}{2019}\natexlab{}.
\newblock \showarticletitle{{NL2Type}: {I}nferring {JavaScript} function types
  from natural language information}. In \bibinfo{booktitle}{\emph{Proceedings
  of the 41st International Conference on Software Engineering, {ICSE} 2019,
  Montreal, QC, Canada, May 25-31, 2019}}. \bibinfo{pages}{304--315}.
\newblock
\urldef\tempurl%
\url{https://doi.org/10.1109/ICSE.2019.00045}
\showDOI{\tempurl}


\bibitem[Mikolov et~al\mbox{.}(2013)]%
        {Mikolov2013}
\bibfield{author}{\bibinfo{person}{Tomas Mikolov}, \bibinfo{person}{Kai Chen},
  \bibinfo{person}{Greg Corrado}, {and} \bibinfo{person}{Jeffrey Dean}.}
  \bibinfo{year}{2013}\natexlab{}.
\newblock \showarticletitle{Efficient Estimation of Word Representations in
  Vector Space}.
\newblock \bibinfo{journal}{\emph{CoRR}}  \bibinfo{volume}{abs/1301.3781}
  (\bibinfo{year}{2013}).
\newblock
\urldef\tempurl%
\url{http://arxiv.org/abs/1301.3781}
\showURL{%
\tempurl}


\bibitem[Patra and Pradel(2021)]%
        {fse2021}
\bibfield{author}{\bibinfo{person}{Jibesh Patra} {and} \bibinfo{person}{Michael
  Pradel}.} \bibinfo{year}{2021}\natexlab{}.
\newblock \showarticletitle{Semantic bug seeding: a learning-based approach for
  creating realistic bugs}. In \bibinfo{booktitle}{\emph{{ESEC/FSE} '21: 29th
  {ACM} Joint European Software Engineering Conference and Symposium on the
  Foundations of Software Engineering, Athens, Greece, August 23-28, 2021}},
  \bibfield{editor}{\bibinfo{person}{Diomidis Spinellis},
  \bibinfo{person}{Georgios Gousios}, \bibinfo{person}{Marsha Chechik}, {and}
  \bibinfo{person}{Massimiliano~Di Penta}} (Eds.). \bibinfo{publisher}{{ACM}},
  \bibinfo{pages}{906--918}.
\newblock
\urldef\tempurl%
\url{https://doi.org/10.1145/3468264.3468623}
\showDOI{\tempurl}


\bibitem[Patra and Pradel(2022)]%
        {icse2022-Nalin}
\bibfield{author}{\bibinfo{person}{Jibesh Patra} {and} \bibinfo{person}{Michael
  Pradel}.} \bibinfo{year}{2022}\natexlab{}.
\newblock \showarticletitle{Nalin: Learning from Runtime Behavior to Find
  Name-Value Inconsistencies in Jupyter Notebooks}. In
  \bibinfo{booktitle}{\emph{ICSE}}.
\newblock


\bibitem[Pennington et~al\mbox{.}(2014)]%
        {glove}
\bibfield{author}{\bibinfo{person}{Jeffrey Pennington},
  \bibinfo{person}{Richard Socher}, {and} \bibinfo{person}{Christopher~D.
  Manning}.} \bibinfo{year}{2014}\natexlab{}.
\newblock \showarticletitle{{GloVe: Global Vectors for Word Representation}}.
  In \bibinfo{booktitle}{\emph{Proceedings of the 2014 Conference on Empirical
  Methods in Natural Language Processing, {EMNLP} 2014, October 25-29, 2014,
  Doha, Qatar, {A} meeting of SIGDAT, a Special Interest Group of the {ACL}}},
  \bibfield{editor}{\bibinfo{person}{Alessandro Moschitti},
  \bibinfo{person}{Bo~Pang}, {and} \bibinfo{person}{Walter Daelemans}} (Eds.).
  \bibinfo{publisher}{{ACL}}, \bibinfo{pages}{1532--1543}.
\newblock
\urldef\tempurl%
\url{https://doi.org/10.3115/v1/d14-1162}
\showDOI{\tempurl}


\bibitem[Pradel and Chandra(2022)]%
        {NeuralSoftwareAnalysis}
\bibfield{author}{\bibinfo{person}{Michael Pradel} {and}
  \bibinfo{person}{Satish Chandra}.} \bibinfo{year}{2022}\natexlab{}.
\newblock \showarticletitle{Neural software analysis}.
\newblock \bibinfo{journal}{\emph{Commun. {ACM}}} \bibinfo{volume}{65},
  \bibinfo{number}{1} (\bibinfo{year}{2022}), \bibinfo{pages}{86--96}.
\newblock
\urldef\tempurl%
\url{https://doi.org/10.1145/3460348}
\showDOI{\tempurl}


\bibitem[Pradel et~al\mbox{.}(2020)]%
        {fse2020}
\bibfield{author}{\bibinfo{person}{Michael Pradel}, \bibinfo{person}{Georgios
  Gousios}, \bibinfo{person}{Jason Liu}, {and} \bibinfo{person}{Satish
  Chandra}.} \bibinfo{year}{2020}\natexlab{}.
\newblock \showarticletitle{TypeWriter: Neural Type Prediction with
  Search-based Validation}. In \bibinfo{booktitle}{\emph{{ESEC/FSE} '20: 28th
  {ACM} Joint European Software Engineering Conference and Symposium on the
  Foundations of Software Engineering, Virtual Event, USA, November 8-13,
  2020}}. \bibinfo{pages}{209--220}.
\newblock
\urldef\tempurl%
\url{https://doi.org/10.1145/3368089.3409715}
\showURL{%
\tempurl}


\bibitem[Pradel and Sen(2018)]%
        {oopsla2018-DeepBugs}
\bibfield{author}{\bibinfo{person}{Michael Pradel} {and}
  \bibinfo{person}{Koushik Sen}.} \bibinfo{year}{2018}\natexlab{}.
\newblock \showarticletitle{{DeepBugs}: A learning approach to name-based bug
  detection}.
\newblock \bibinfo{journal}{\emph{{PACMPL}}} \bibinfo{volume}{2},
  \bibinfo{number}{{OOPSLA}} (\bibinfo{year}{2018}),
  \bibinfo{pages}{147:1--147:25}.
\newblock
\urldef\tempurl%
\url{https://doi.org/10.1145/3276517}
\showURL{%
\tempurl}


\bibitem[Rasthofer et~al\mbox{.}(2014)]%
        {susi}
\bibfield{author}{\bibinfo{person}{Siegfried Rasthofer},
  \bibinfo{person}{Steven Arzt}, {and} \bibinfo{person}{Eric Bodden}.}
  \bibinfo{year}{2014}\natexlab{}.
\newblock \showarticletitle{A Machine-learning Approach for Classifying and
  Categorizing Android Sources and Sinks}. In \bibinfo{booktitle}{\emph{21st
  Annual Network and Distributed System Security Symposium, {NDSS} 2014, San
  Diego, California, USA, February 23-26, 2014}}. \bibinfo{publisher}{The
  Internet Society}.
\newblock
\urldef\tempurl%
\url{https://www.ndss-symposium.org/ndss2014/machine-learning-approach-classifying-and-categorizing-android-sources-and-sinks}
\showURL{%
\tempurl}


\bibitem[Sch{\"o}lkopf et~al\mbox{.}(2001)]%
        {scholkopf2001estimating}
\bibfield{author}{\bibinfo{person}{Bernhard Sch{\"o}lkopf},
  \bibinfo{person}{John~C Platt}, \bibinfo{person}{John Shawe-Taylor},
  \bibinfo{person}{Alex~J Smola}, {and} \bibinfo{person}{Robert~C Williamson}.}
  \bibinfo{year}{2001}\natexlab{}.
\newblock \showarticletitle{Estimating the support of a high-dimensional
  distribution}.
\newblock \bibinfo{journal}{\emph{Neural computation}} \bibinfo{volume}{13},
  \bibinfo{number}{7} (\bibinfo{year}{2001}), \bibinfo{pages}{1443--1471}.
\newblock


\bibitem[She et~al\mbox{.}(2019)]%
        {She2019}
\bibfield{author}{\bibinfo{person}{Dongdong She}, \bibinfo{person}{Yizheng
  Chen}, \bibinfo{person}{Baishakhi Ray}, {and} \bibinfo{person}{Suman Jana}.}
  \bibinfo{year}{2019}\natexlab{}.
\newblock \showarticletitle{Neutaint: Efficient Dynamic Taint Analysis with
  Neural Networks}.
\newblock \bibinfo{journal}{\emph{CoRR}}  \bibinfo{volume}{abs/1907.03756}
  (\bibinfo{year}{2019}).
\newblock
\showeprint[arxiv]{1907.03756}
\urldef\tempurl%
\url{http://arxiv.org/abs/1907.03756}
\showURL{%
\tempurl}


\bibitem[Utture et~al\mbox{.}(2022)]%
        {Utture2022}
\bibfield{author}{\bibinfo{person}{Akshay Utture}, \bibinfo{person}{Shuyang
  Liu}, \bibinfo{person}{Christian~Gram Kalhauge}, {and} \bibinfo{person}{Jens
  Palsberg}.} \bibinfo{year}{2022}\natexlab{}.
\newblock \showarticletitle{Striking a Balance: Pruning False-Positives from
  Static Call Graphs}. In \bibinfo{booktitle}{\emph{ICSE}}.
\newblock


\bibitem[Vasic et~al\mbox{.}(2019)]%
        {Vasic2019}
\bibfield{author}{\bibinfo{person}{Marko Vasic}, \bibinfo{person}{Aditya
  Kanade}, \bibinfo{person}{Petros Maniatis}, \bibinfo{person}{David Bieber},
  {and} \bibinfo{person}{Rishabh Singh}.} \bibinfo{year}{2019}\natexlab{}.
\newblock \showarticletitle{Neural Program Repair by Jointly Learning to
  Localize and Repair}. In \bibinfo{booktitle}{\emph{ICLR}}.
\newblock


\bibitem[Wainakh et~al\mbox{.}(2021)]%
        {icse2021}
\bibfield{author}{\bibinfo{person}{Yaza Wainakh}, \bibinfo{person}{Moiz Rauf},
  {and} \bibinfo{person}{Michael Pradel}.} \bibinfo{year}{2021}\natexlab{}.
\newblock \showarticletitle{IdBench: Evaluating Semantic Representations of
  Identifier Names in Source Code}. In \bibinfo{booktitle}{\emph{43rd
  {IEEE/ACM} International Conference on Software Engineering, {ICSE} 2021,
  Madrid, Spain, 22-30 May 2021}}. \bibinfo{publisher}{{IEEE}},
  \bibinfo{pages}{562--573}.
\newblock
\urldef\tempurl%
\url{https://doi.org/10.1109/ICSE43902.2021.00059}
\showDOI{\tempurl}


\bibitem[Ye et~al\mbox{.}(2022)]%
        {Ye2022a}
\bibfield{author}{\bibinfo{person}{He Ye}, \bibinfo{person}{Matias Martinez},
  {and} \bibinfo{person}{Martin Monperrus}.} \bibinfo{year}{2022}\natexlab{}.
\newblock \showarticletitle{Neural Program Repair with Execution-based
  Backpropagation}. In \bibinfo{booktitle}{\emph{ICSE}}.
\newblock


\end{thebibliography}

\end{document}